\documentclass[11pt,a4paper]{article}
\usepackage[utf8]{inputenc}
\usepackage[english]{babel}
\usepackage{amsmath}
\usepackage{amsfonts}
\usepackage{amssymb}
\usepackage{graphicx}
\usepackage[left=2cm,right=2cm,top=2cm,bottom=2cm]{geometry}

\usepackage{bm}
\usepackage{bbm}
\usepackage{amsmath,amsthm,amssymb} 
\usepackage{pifont}
\usepackage{url}

\usepackage{graphicx}
\usepackage{subfigure}

\usepackage{cite}
\usepackage{cleveref}

\usepackage[all,pdf]{xy}

\usepackage{booktabs}
\usepackage{multirow}
\usepackage{multicol}
\usepackage{stfloats}
\usepackage{diagbox}

\DeclareMathOperator{\dif}{d}

\newcommand{\mpair}[2]{ \left\langle {#1}, {#2} \right\rangle}

\newcommand{\me} {\mathrm{e}}

\newcommand{\fracode}[2]{\frac{\dif {#1}}{\dif {#2}}}         

\newcommand{\set}[1]{\left\{ #1 \right\}}
\newcommand{\abs}[1]{\left| #1 \right|}

\newcommand{\normp}[2]{{\left\lVert #1 \right\rVert}_{#2}}

\newcommand{\inv}[1]{#1^{-1}}

\newcommand{\ES}[3]{\mathbb{#1}^{{#2}\times {#3}}}     

\newcommand{\scrd}[2]{{#1}_{\mathrm{#2}}}

\newcommand{\BigO}[1]{\mathcal{O}\left(#1\right)}

\usepackage{cleveref}
\usepackage{algorithm}
\usepackage{algorithmicx}
\usepackage{algpseudocode}
\usepackage{float}

\usepackage{forest}

\usepackage{longtable}
\usepackage{supertabular}
\usepackage{caption}

\newcommand{\Algor}{\textbf{Algorithm}~}
\newcommand{\Fig}{\textbf{Figure}~}
\newcommand{\Tab}{\textbf{Table}~}

\algnewcommand\algorithmicswitch{\textbf{switch}}
\algnewcommand\algorithmiccase{\textbf{case}}
\algnewcommand\algorithmicdefault{\textbf{default}}
\algnewcommand\algorithmicassert{\cpvar{assert}}
\algnewcommand\Assert[1]{\State \algorithmicassert(#1)}%
\algdef{SE}[SWITCH]{Switch}{EndSwitch}[1]{\algorithmicswitch\ #1\ \algorithmicdo}{\algorithmicend\ \algorithmicswitch}%
\algdef{SE}[CASE]{Case}{EndCase}[1]{\algorithmiccase\ #1}{\algorithmicend\ \algorithmiccase}%
\algdef{SE}[DEFAULT]{Default}{EndDefault}[1]{\algorithmicdefault\ #1}{\algorithmicend\ \algorithmicdefault}%

\makeatletter
\renewcommand{\ALG@name}{Algorithm}
\newenvironment{breakablealgorithm}
{
	\begin{center}
		\refstepcounter{algorithm}
		\setlength{\baselineskip}{15pt}
		\renewcommand{\caption}[2][\relax]{
			\hrule height.9pt depth0pt \kern3pt
			{\raggedright\textbf{\ALG@name~\thealgorithm} ##2\par}%
			\ifx\relax##1\relax 
			\addcontentsline{loa}{algorithm}{
				\protect\numberline{\thealgorithm}##2}%
			\else 
			\addcontentsline{loa}{algorithm}{
				\protect\numberline{\thealgorithm}##1}%
			\fi
			\kern2pt\hrule\kern2pt
		}
	}{
		\kern3pt\hrule\relax
	\end{center}
}

\usepackage{caption}
\usepackage{listings}
\newcommand{\cpvar}[1]{\texttt{#1}}

\newcommand{\ProcName}[1]{\textsc{#1}}

\author{Hong-Yan Zhang\thanks{Corresponding author: Hong-Yan Zhang;  email: hongyan@hainnu.edu.cn; ORCID: 0000-0002-4400-9133}, Wei Sun, Xiao Chen,  Rui-Jia Lin and Yu Zhou\\ \quad \\ \small{School of Information Science and Technology, Hainan Normal University, Haikou 571158, China}}

\title{\textbf{Fixed-Point Algorithm for Solving the  Critical Value and Upper Tail Quantile of Kuiper's Statistics}}
\begin{document}
\maketitle

\begin{abstract}
Kuiper's statistic is a good measure for the difference of ideal distribution and empirical distribution  in the goodness-of-fit test. However, it is a challenging problem to  solve  the critical value and upper tail quantile, or simply Kuiper pair, of Kuiper's statistics due to the difficulties of solving the nonlinear equation and reasonable approximation of infinite series. In this work, the contributions lie in three perspectives: firstly, the second order approximation for the  infinite series of the cumulative distribution of the critical value is used to achieve higher precision; secondly, the principles and fixed-point algorithms for solving the Kuiper pair are presented with details; finally, finally, a mistake about the critical value $c^\alpha_n$ for $(\alpha, n)=(0.01,30)$ in Kuiper's distribution table has been labeled and corrected where $n$ is the sample capacity and $\alpha$ is the upper tail quantile.  The algorithms are verified and validated by comparing with the table provided by Kuiper. The methods and algorithms proposed are enlightening and worth of introducing to the college students, computer programmers, engineers, experimental psychologists and so on.  
 \\
\noindent \textbf{Kerwords}: Kuiper's statistic; Upper tail quantile; Fixed-point;
Numerical approximation; Algorithm Design; STEM education
\end{abstract}

\tableofcontents


\section{Introduction}

In statistics, for two cumulative distribution functions (CDF) $F_1(x)$ and $F_2(x)$, Kuiper's test 
\cite{Kuiper1960TestsCR,Stephens1970,Pearson1972,KStest2008,KnuthTAOCP2,
Jin2015,Press1992KuiperTest,
Koch2020Benfords,Dowd2020ECDF,Lanzante2021} is defined by
\begin{equation}
V = D^+ + D^- = \sup_x [F_1(x) - F_2(x)] + \sup_x[F_2(x) - F_1(x)]
\end{equation}
where the discrepancy statistics 
\begin{equation}
D^+ = \sup_{x\in \mathbb{R}} [F_1(x) - F_2(x)], \quad D^- = \sup_{x\in \mathbb{R}} [F_2(x) - F_1(x)] = -\inf_{x\in \mathbb{R}}[F_1(x) - F_2(x)]
\end{equation}
represent the maximum deviation above and below the two cumulative distribution functions being 
compared, respectively. The trick with Kuiper's test is to use the quantity $ D^+ + D^-$  
as the test statistic instead of $\displaystyle \sup_x\abs{F_1(x) - F_2(x)}$ in Kolmogrov-Smirnov test. 
This small change makes Kuiper's test as sensitive in the tails as at the median and 
also makes it invariant under cyclic transformations of the independent variable. 
The Kuiper's test is useful for the goodness-of-fit test, also named by distribution fitting test. 
Exactly speaking, there are two types of Kuiper's test \cite{Kuiper1960TestsCR} 
\begin{itemize}
\item $V_n$-test, which is used for comparing two distributions with $n$ samples;  
\item $V_{n,m}$-test, which is used for comparing two distributions with $n$ and $m$  samples respectively.
\end{itemize}

Suppose that $F_{V_n}(x)$ is the CDF of the Kuiper's statistic $V_n$, $\alpha$ is the \textit{upper tail probability}\footnote{The upper tail probability will be the upper tail significance level or the probability of first type error in hypothesis test.} for the Kuiper's test statistic and $v^\alpha_n$ is the \textit{upper tail quantile} (UTQ,  also named \textit{upper tail fractile})  for $V_n$, then
\begin{equation} \label{eq-alpha-def}
\alpha = \Pr\set{V_n >v^\alpha_n} = 1- \Pr\set{V_n \le v^\alpha_n} = 1- F_{V_n}(v^\alpha_n).
\end{equation}
Formally, we have
\begin{equation} \label{eq-significance-level}
v^\alpha_n = \inv{F}_{V_n}(1-\alpha).
\end{equation}
For $\forall \alpha\in [0,1]$,  we always have
\begin{equation}
v_{1-\alpha}^n = v^\alpha_n
\end{equation}
where $v_{1-\alpha}^n$ is the lower tail quantile for the lower tail probability $1-\alpha$.  Therefore, 
\begin{equation} \label{eq-inv-CDF}
v_\alpha^n = v^{1-\alpha}_n = \inv{F}_{V_n}(\alpha), \quad \forall \alpha\in [0, 1],
\end{equation}
which implies that \textit{solving the lower or upper tail probability is equivalent to solve the inverse of the CDF}. 
Unfortunately, there is  no simple expression for $F_{V_n}(\cdot)$ and it is difficult to calculate the inverse $\inv{F}_{V_n}(\cdot)$. 

Moreover, for sufficiently large $n$, the statistic $V_n = D^+_n + D^-_n$ will approach to $0$, which implies that the curve of $F_{V_n}(x)$ will become more and more steep with the increasing of $n$ and 
\begin{equation} \label{eq-lim-Vn-CDF}
\lim_{n\to \infty} F_{V_n}(x) = 
\left\{
\begin{aligned}
1, & & x \ge 0;\\
0, & & x < 0.
\end{aligned}
\right.
\end{equation}

Although the Kuiper's test has been proposed for about 50 years, it is not widely known to college students, computer programmers, engineers, experimental psychologist and so on partly due to the difficulty of solving the upper tail quantile and partly due to the lack of open software for automatic calculation. 
Let 
\begin{equation}
K_n = \sqrt{n}\cdot V_n = \sqrt{n}\cdot (D^+_n + D^-_n)
\end{equation}
be the Kuiper's $K_n$ statistic of critical value for the $V_n$-test, Kuiper \cite{Kuiper1960TestsCR,Darling1957} pointed out that 
\begin{equation} \label{eq-utp-inf-series}
\begin{aligned}
\Pr\set{K_n \le c} 
= 1-\sum^\infty_{j=1}2(4j^2c^2-1)\me^{-2j^2c^2} 
  + \frac{8}{3\sqrt{n}}c\sum^\infty_{j=1}j^2(4j^2c^2-3)\me^{-2j^2c^2} + \BigO{\frac{1}{n}},
\end{aligned}
\end{equation}
for the positive critical value $c$. It should be noted that the upper bound for the  approximation error is $\BigO{n^{-1}}$, which implies this formula is appropriate for large sample capacity $n
$.  Kuiper' s approximation
\begin{equation}
\alpha =  \Pr\set{K_n > c} \approx \left[-2 + \frac{8}{\sqrt{n}}c + 8c^2 -\frac{32}{3\sqrt{n}}c^3 \right]\me^{-2c^2}, \quad c > \frac{6}{5},
\end{equation}
 for solving $c$ is based on the terms in which $j=1$ of the two infinite series in \eqref{eq-utp-inf-series}. 
There are some disadvantages for this approximation: 
\begin{itemize}
\item the terms in which $j=2$ in the two infinite series have been ignored, which not only reduces the numerical precision of the solution $c$ due to the term $\BigO{n^{-1}}$ for small $n$ but also reduces the feasible 
region for the initial value of $c$, denoted by $\scrd{c}{guess}$, in numerical computation; 
\item the condition $c > 6/5$ is not appropriate and smaller lower bound for $c$ is allowed. 
\end{itemize}
Stephens \cite{Stephens1970} proposed the modified statistic for the replacement of $V_n$, but his approximation   
\begin{equation}
\alpha = (8c^2-2)\me^{-2c^2},
\end{equation} 
is based on the first term of the first infinite series in \eqref{eq-alpha-approx}. Obviously, this approximation does not depend on the parameter $n$ and its precision will be affected significantly.
In other words, this approximation just works well for sufficiently large $n$, which   means that the number of random samples should be large enough in the sense of data analysis. 
Furthermore, Kuiper discussed the $V_{n,n}$-test but Stephens ignored it. 
It is a common dark side of both Kuiper's work and Stephens's work that there are lack of
numerical algorithms for solving the numerical solution to the critical value $c$ for the given upper tail
probability $\alpha$ in the era of 1960s and 1970s when only a few people can use computers and the computer science was not fully developed.

It is easy to find that the two infinite sequences in \eqref{eq-utp-inf-series} converge rapidly and 
it is a better approximation if we consider the first two terms when compared with the schemes taken by
Kuiper and Stephens. By taking the first two terms for the infinite series, we can deduce that
\begin{equation} \label{eq-alpha-approx}
\begin{aligned}
 \alpha & = \Pr\set{K_n > c} 
=  \sum^\infty_{j=1}2(4j^2c^2-1)\me^{-2j^2c^2} 
   - \frac{8}{3\sqrt{n}}c\sum^\infty_{j=1}j^2(4j^2c^2-3)\me^{-2j^2c^2} + \BigO{\frac{1}{n}}\\
&\approx  \left[ 2(4c^2-1) - \frac{8}{3\sqrt{n}}c(4c^2-3)\right]\me^{-2c^2} 
  +\left[ 2(16c^2-1) -\frac{32}{3\sqrt{n}}c(16c^2-3)\right]\me^{-8c^2}\\
&= \left[ -2 + \frac{8}{\sqrt{n}}c + 8c^2 -\frac{32}{3\sqrt{n}}c^3 \right]\me^{-2c^2} 
  +\left[ -2 + \frac{32}{\sqrt{n}}c + 32c^2 -\frac{512}{3\sqrt{n}}c^3 \right]\me^{-8c^2}.\\
\end{aligned}
\end{equation}
This approximation is a second order approximation for infinite series about the index $j$, which is more reasonable due to the term $\BigO{n^{-1}}$ in  \eqref{eq-alpha-approx}.

 In this paper, our emphasis is put on the numerical algorithms for automatic approaches for solving the upper tail critical value and quantile of Kuiper's statistic. The contents of this paper are organized as follows: Section \ref{sec-preliminary} gives the preliminaries about the principle and algorithm for fixed-point equation; Section \ref{sec-comp-method-Kuiper-Vn} focuses on the computational method for solving Kuiper's pairs of $V_n$-test and $V_{n,n}$-test; Section
\ref{sec-algorithms} presents the algorithms with pseudo-code; Section \ref{sec-V-V} demonstrates the numerical results which is comparable with Kuiper's original results; Section \ref{sec-discussion} discussed the impact of the choices of nonlinear equation and initial value on the numerical solution; and finally 
Section \ref{sec-conclusion} is the conclusion.

\section{Preliminaries} \label{sec-preliminary}

\subsection{Principle of Iterative Method for Fixed-Point}

Although there are various fixed-point theorems \cite{Zeidler1995AFAvol1,Atkinson2009} for different situation and applications, we just consider the fixed-point iterative method for solving nonlinear equations according to our objective of solving Kuiper's pairs $\mpair{c^\alpha_n}{v^\alpha_n}$ and $\mpair{c^\alpha_{n,n}}{v^\alpha_{n,n}}$.

For a nonlinear equation
\begin{equation}
f(x) = 0, \quad x\in [a,b]
\end{equation}
usually we can convert it into a fixed-point equation equivalently
\begin{equation}
x = T(f, x), \quad x\in[a,b]
\end{equation}
where $f(\cdot)$ is a mapping and $T(\cdot, \cdot)$ is a  contractive operator. There are two typical categories of fixed-point equation:
\begin{itemize}
\item[(i)] Direct iterative scheme
      \begin{equation}
         x_{i+1} = T(\scrd{f}{ctm}(x_i), x_i) = \scrd{f}{ctm}(x_i)
      \end{equation}
      which depends on the function $\scrd{f}{ctm}$.
\item[(ii)] Newton's iterative scheme
      \begin{equation}
      x_{i+1} = T(\scrd{f}{nlm}(x_i), x_i) = x_i - \frac{\scrd{f}{nlm}(x_i)}{\scrd{f}{nlm}'(x_i)},
      \end{equation}
      which depends not only on the equivalent nonlinear mapping  
$\scrd{f}{nlm}(x)$ from
\begin{equation}
\phi(x) = 0 \Longleftrightarrow \scrd{f}{nlm}(x) = 0
\end{equation}
but also on its derivative $\scrd{f}{nlm}'(x)$.
\end{itemize}
We remark that the strings "ctm" and "nlm" in the subscripts  means \underline{c}on\underline{t}ractive \underline{m}apping and \underline{n}on\underline{l}inear \underline{m}apping respectively.

The initial value, say $\scrd{x}{guess}$, should be set carefully since there is a domain for the contractive mapping $T$ generally. For mathematicians, it is necessary to prove the feasibility of choice 
$\scrd{x}{guess}$. For engineers and computer programmers, they may prefer to draw the curve of $ y = x  - T(f,x)$ for $x\in [a, b]$ and observe the interval for of the root to set the initial value $\scrd{x}{guess}$, which is more intuitive and efficient than rigorous mathematical analysis.

The stopping condition for the iterative method is satisfied by the Cauchy's criteria for the convergent sequence $\set{x_i: i = 0, 1, 2, 3, \cdots}$
\begin{equation}
d(x_{i+1}, x_i) < \epsilon
\end{equation}
where $\epsilon$ denotes the precision and $d(\cdot,\cdot)$ is a metric for the difference of the current value of $x$, say $\scrd{x}{guess}$ and its updated version, say $\scrd{x}{improve}$. The simplest choice of $d(\cdot, \cdot)$ is the absolute value function for $x\in [a, b] \in \mathbb{R}$, i.e.,
\begin{equation}
d(x_i, x_{i+1}) = d(x_{i+1}, x_i) = \abs{x_{i+1} - x_i}
\end{equation}

For practical problems arising in various fields, there may be some extra parameters in the function $\phi$, say $\phi(x, \alpha, n)$ where $\alpha$ is a real number, $n$ is an integer, where the order of $x, \alpha, n$ is not important\footnote{In other words, both $\phi(x,\alpha,n)$ and $\phi(x,n,\alpha)$ are feasible and correct for computer programming.}. In general, we can use the notations $\phi(x, \mu_1, \cdots, \mu_r), f(x, \mu_1 \cdots \mu_r)$ and $T(f, x, \mu_1, \cdots, \mu_r)$ to represent the scenario when there are some extra parameters. We remark that
the derivative of $f(x,\mu_1, \cdots, m_r)$ with respect to the variable $x$ should be calculated by
\begin{equation}
     f'(x, \mu_1, \cdots \mu_r) = \lim_{\Delta x\to 0} \frac{f(x+\Delta x, \mu_1, \cdots, \mu_r) -f(x,\mu_1, \cdots, \mu_r)}{\Delta x}
\end{equation}
for the Newton's iterative method.

It should be remarked the abstraction level is essential for designing feasible algorithm to solve the fixed-point of the form
\begin{equation} \label{eq-fixedpoint}
x = T(f, x, \mu_1, \cdots, \mu_r) = \mathcal{T}^*_\star(f, x), \quad x\in [a,b]
\end{equation}
for the contractive mapping $T$ or its equivalent form $\mathcal{T}^*_\star$, which is called the \textit{updating function} in programming language. Note that the extra
parameters $\mu_1, \cdots, \mu_r$ are moved to upper subscript $*$ and subscript $\star$ such that
 $(*, \star)=(\mu_1, \cdots, \mu_r)$  in order to emphasize the key variable $x$ and function $f$. In the sense of functional analysis, the $T$ or $\mathcal{T}^*_\star$ in  \eqref{eq-fixedpoint} is an abstract function. By comparison, in the sense of programming language in computer science, the $T$ or $\mathcal{T}^*_\star$ is a \textit{high order function} and $f$ is a function object which is called a \textit{pointer to function} in C/C++98, or a \textit{$\lambda$-expression} in Lisp/Haskell/Julia/Python/C++11/Java, or \textit{function handle} in Octave/MATLAB, and so on. 

\subsection{Unified Framework and Algorithm for Solving Fixed-point}

We now give the pseudo-code for the fixed-point algorithms with the concepts of high order function and function object, please see  \Algor \ref{alg-abstract-fixed-point}. Note that the order for the list of arguments can be set according to the programmer's preferences. If a default value for the $\scrd{x}{guess}$  should  be configured, it is wise to set the $\scrd{x}{guess}$ as the last formal argument for the convenience of implementation with some concrete computer programming language such as the C++.

\begin{breakablealgorithm}
\caption{Unified Framework for Solving the Fixed-Point of $x = \mathcal{T}^*_\star(f, x)= T(f, x, \mu_1, \cdots, \mu_r)$}\label{alg-abstract-fixed-point}
\begin{algorithmic}[1]
\Require Contractive mapping  $T$ as the updator which is a high order function, function object $f$ for the $\scrd{f}{nlm}$ or $\scrd{f}{ctm}$, function object $d$ for the distance $d(x_i, x_{i+1})$, precision $\epsilon$, initial value $\scrd{x}{guess}$ and extra parameters $\mu_1, \cdots, \mu_r$ with the same or different data types.
\Ensure Fixed-point $x$ such that $x = \mathcal{T}^*_\star(f, x) = T(f, x, \mu_1, \cdots, \mu_r)$
\Function{FixedPointSolver}{$T, f, d, \epsilon, \scrd{x}{guess},  \mu_1, \cdots, \mu_r$}
\State $\scrd{x}{improve}\gets T(f, \scrd{x}{guess}, \mu_1, \cdots, \mu_r)$;
\While{$d(\scrd{x}{improve}, \scrd{x}{guess}) \ge \epsilon$}
\State $\scrd{x}{guess} \gets \scrd{x}{improve}$;
\State $\scrd{x}{improve}\gets T(f, \scrd{x}{guess}, \mu_1, \cdots, \mu_r)$;
\EndWhile
\State \Return $\scrd{x}{improve}$;
\EndFunction
\end{algorithmic}
\end{breakablealgorithm}

In the sense of programming language and discrete mathematics, $f$ is an ordinary (first order) function, $T$ is a second order function and $\ProcName{FixedPointSolve}$ is a third order function. 

\begin{breakablealgorithm}
\caption{Calculate the distance of $x$ and $y$}
\label{alg-calc-dist}
\begin{algorithmic}[1]
\Require $x, y\in \mathbb{R}$
\Ensure The distance of $x$ and $y$, i.e., $d(x,y)$ 
\Function{Distance}{$x,y$}
\State $\cpvar{dist} \gets \abs{x -y}$; \quad // if $x, y\in \ES{R}{n}{1}$, we can use the $\ell^p$-norm to get the distance by $\normp{x-y}{p}$
\State \Return $\cpvar{dist}$;
\EndFunction
\end{algorithmic}
\end{breakablealgorithm}

\section{Computational Method of Critical Value and Quantile for Kuiper's Test} 
\label{sec-comp-method-Kuiper-Vn}

In this paper, our key issue here is to design algorithm to compute the $\inv{F}_{V_n}(1-\alpha)$ 
for the given upper tail quantile $\alpha$ and the integer $n$.

\subsection{Computation of Kuiper's Critical Value in $V_n$-test}

Put
\begin{equation} \label{eq-def-A-B}
A_1(c, n) = -2 + \frac{8}{\sqrt{n}}c + 8c^2 -\frac{32}{3\sqrt{n}}c^3, \quad
A_2(c, n) = -2 + \frac{32}{\sqrt{n}}c + 32c^2 -\frac{512}{3\sqrt{n}}c^3 
\end{equation}
and substitute \eqref{eq-def-A-B} into \eqref{eq-alpha-approx}, we can obtain the key equation for specifying the critical value $c$ for the given $\alpha$ and $n$ as follows
\begin{equation}\label{eq-utp-alpha-c}
\boxed{2c^2 + \ln \alpha = \ln\left[A_1(c, n) + A_2(c, n)\cdot \me^{-6c^2}\right]}.
\end{equation}
Let 
\begin{equation} \label{eq-Ganc-def}
\scrd{f}{nlm1}(c, \alpha, n)  = 2c^2 + \ln \alpha - \ln \left[ A_1(c, n)
  +A_2(c, n) \cdot \me^{-6c^2}\right],
\end{equation}
where the digit 1 in the string "nlm1" means single $n$ for the notation $V_n$, 
then the critical value $c_\alpha$ is the solution of the following non-linear equation
\begin{equation} \label{eq-Ganc-root}
\scrd{f}{nlm1}(c, \alpha, n) = 0
\end{equation}
where $\scrd{f}{nlm1}$ is a concrete version of $\phi(x, \mu_1, \cdots, \mu_r) = 0$ with two extra parameters $\mu_1=\alpha$ and $\mu_2=n$.
There are various methods for solving non-linear equation \eqref{eq-Ganc-root} and our method  is based on the fixed-point theorem \cite{Zeidler1995AFAvol1} and iterative algorithm. 

A necessary condition for \eqref{eq-alpha-approx} is that $\alpha = \Pr\set{\sqrt{n}\cdot V_n > c} >0$ for any   $n\in \mathbb{N}$. Let $n\to \infty$, we immediately have
\begin{equation}
A_1(\infty, c) = -2 + 8c^2 > 0, \quad A_2(\infty, c) = -2 + 32c^2 > 0.
\end{equation} 
Thus, we get the following necessary condition
\begin{equation}
c > \frac{1}{2}
\end{equation}
for the function $g(c, \alpha, n)$. Kuiper \cite{Kuiper1960TestsCR} pointed out that a necessary condition for $c$ is $c > 6/5 = 1.2$ when the first term is considered in the infinite series for \eqref{eq-alpha-approx}. On the other hand, the larger the $n$ is, the larger $c$ is feasible according to the definition of
critical value in the sense of large samples in statistics. Hence for fixed $\alpha$ and $n\to \infty$, we can find that $c^\alpha_\infty$ gives the upper bound of the variable $c$. For  $\alpha = 10^{-10}$, we can find that $c^\alpha_\infty \approx 3.7226 < 4$. 

\subsubsection{Direct Iterative Method}

Let
\begin{equation}
\scrd{f}{ctm1}(c,\alpha,n) =\sqrt{\frac{\ln\left[A_1(c, n) + A_2(c, n)\cdot \me^{-6c^2}\right] - \ln \alpha}{2}},
\end{equation}
then \eqref{eq-utp-alpha-c} implies that  
\begin{equation}
 c = \mathcal{A}^\alpha_n(\scrd{f}{ctm1}, c) = \scrd{f}{ctm1}(c, \alpha, n).
\end{equation}
It is easy to verify that $\mathcal{A}^\alpha_n(\scrd{f}{ctm1}, \cdot) $ is a contractive mapping for the given positive integer $n$ and upper tail probability $\alpha\in(0,1)$ for appropriate domain of $c$, say  $c\in (0.5, 2.5)$. Consequently, the
fixed-point theorem implies that the iterative formula
\begin{equation}
c_{i+1} = \mathcal{A}^\alpha_n(\scrd{f}{ctm1}, c_i), \quad i = 0, 1, 2, \cdots
\end{equation}
must converge if the initial value $\scrd{c}{guess}\in (0.5, 2.5)$. 


\begin{figure}[htbp]
\centering
\subfigure[$\alpha=0.10, n = 30, c^\alpha_n = 1.5503$]{
\includegraphics[width=0.38\textwidth]{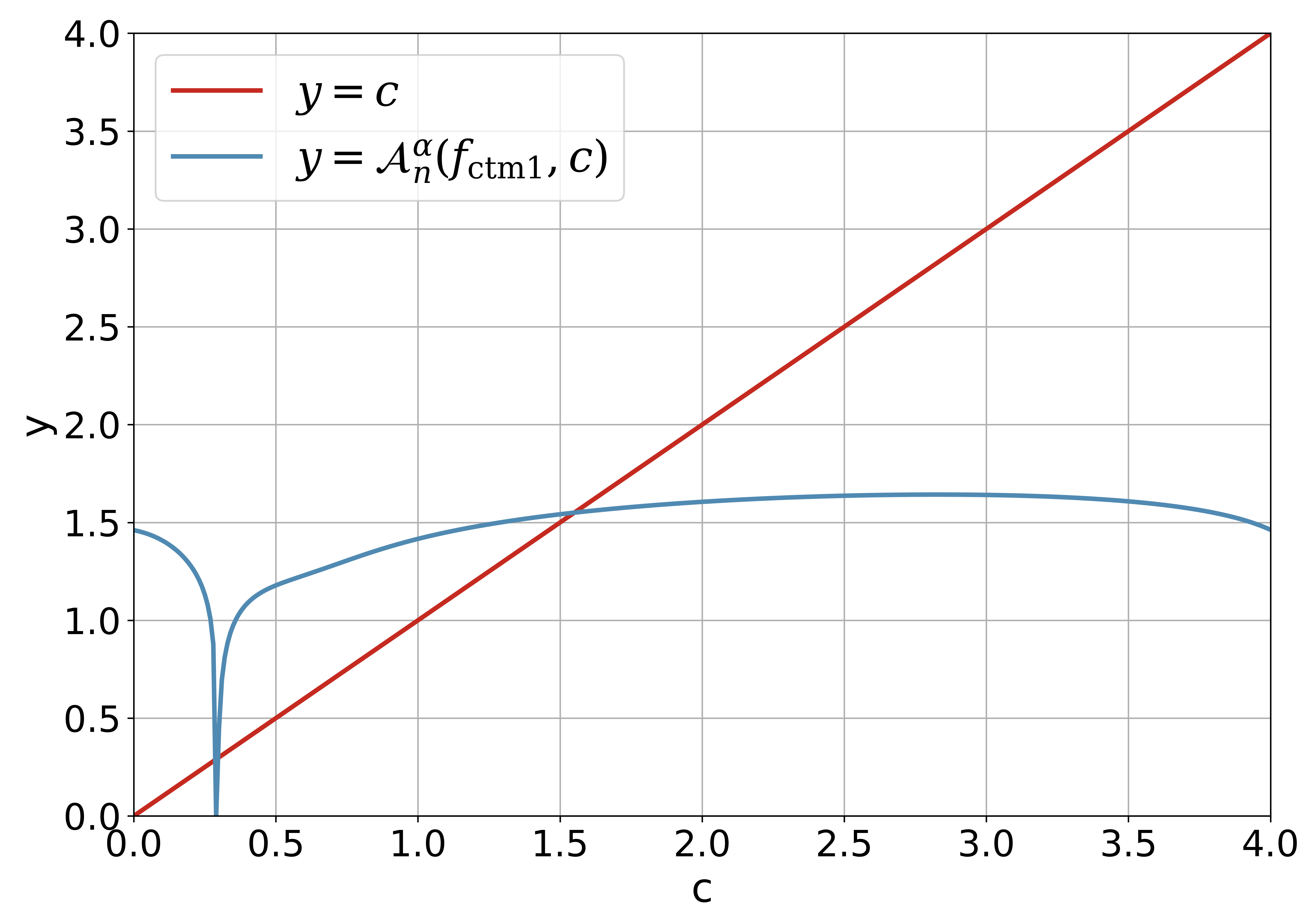} 
} 
\subfigure[$\alpha=0.05, n = 30, c^\alpha_n = 1.6758$]{
\includegraphics[width=0.38\textwidth]{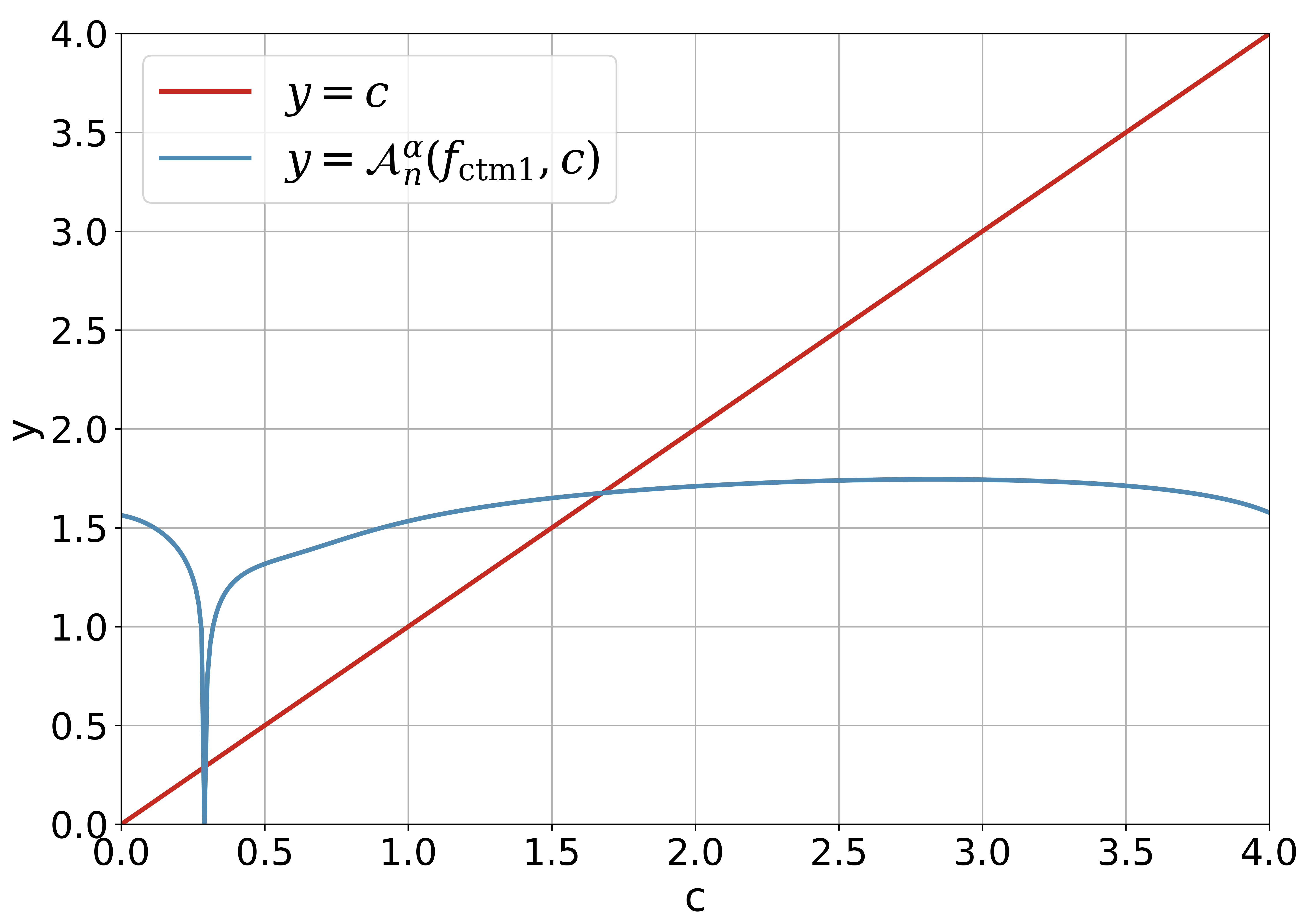} 
}
\subfigure[$\alpha=0.02, n = 30, c^\alpha_n = 1.8235$]{
\includegraphics[width=0.38\textwidth]{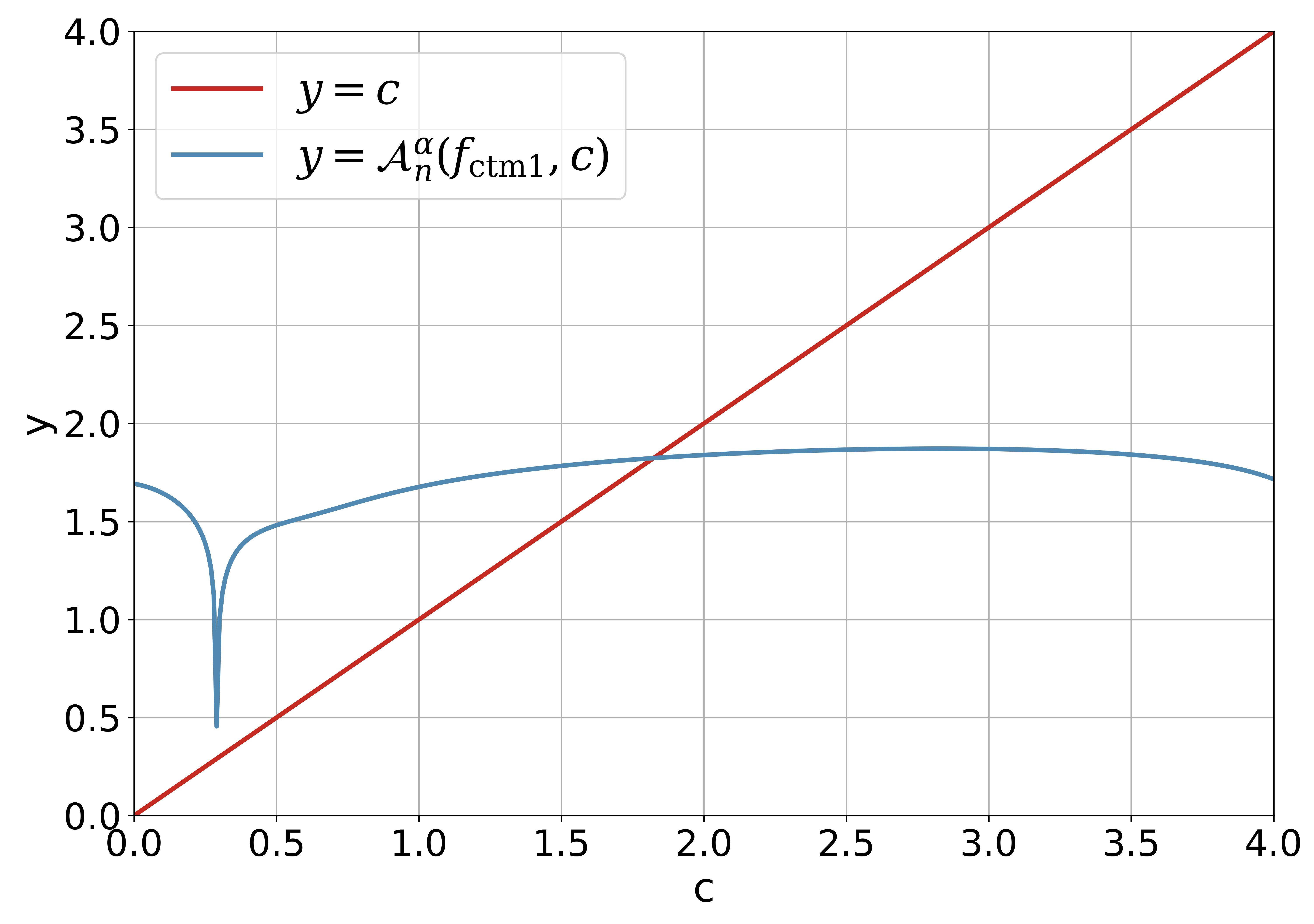} 
}
\subfigure[$\alpha=0.01, n = 30, c^\alpha_n = 1.9252$]{
\includegraphics[width=0.38\textwidth]{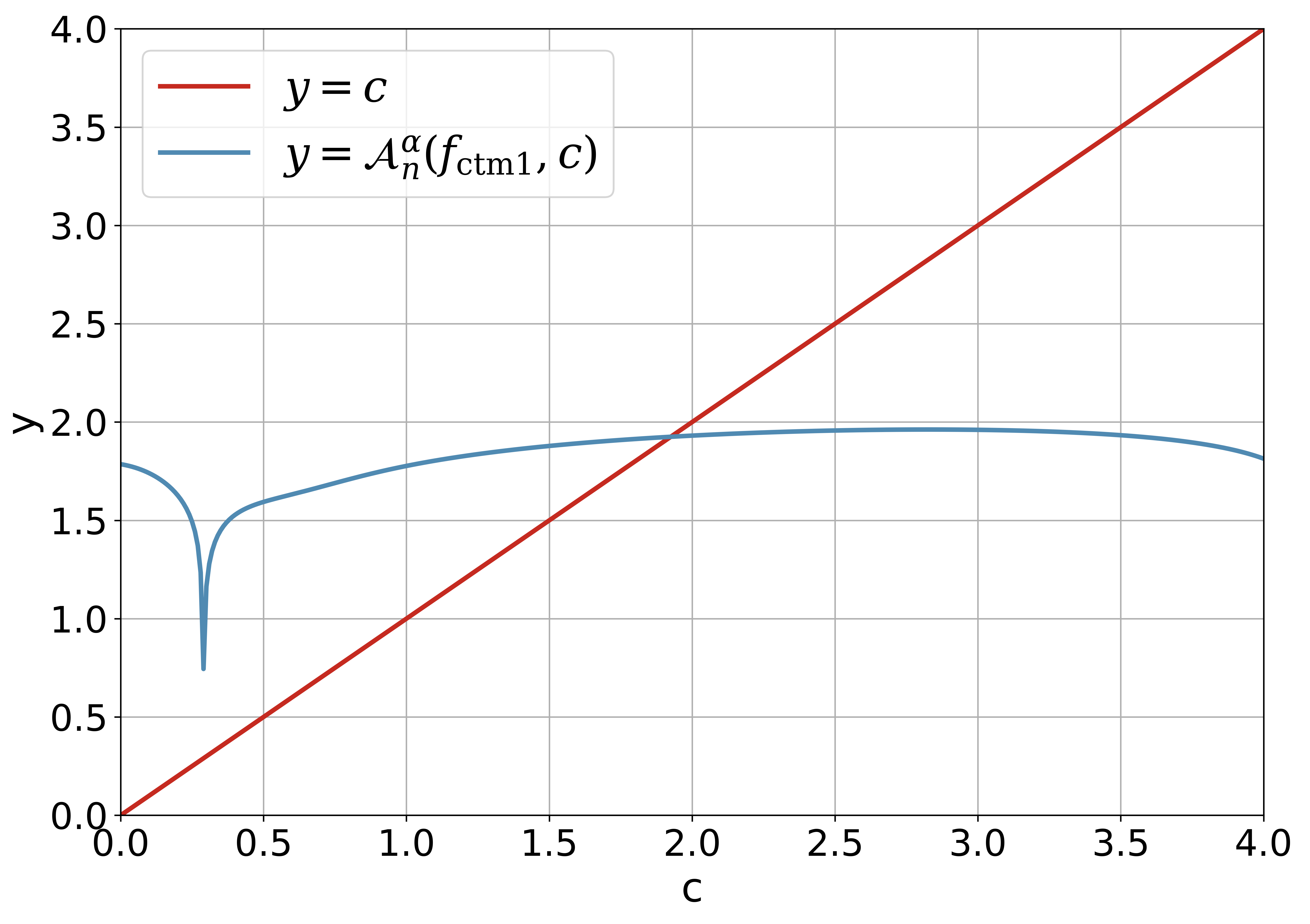} 
}
\caption{The intersection of $ y = c$ and $y = \mathcal{A}^\alpha_n(\scrd{f}{ctm1},c) = \scrd{f}{ctm1}(c,\alpha,n)$ for $\alpha\in \set{0.10, 0.05, 0.02, 0.01}$ and $n=30$.}
\label{fig-Aalphan-c}
\end{figure}

\Fig \ref{fig-Aalphan-c} demonstrates the 
intersection of $y = \mathcal{A}^\alpha_n(\scrd{f}{ctm1},c)$ and $y =c$ clearly. It is easy to find that the absolute value of the derivative $\abs{\fracode{}{c} \mathcal{A}^\alpha_n(\scrd{f}{ctm1},c)} < 1 $ (and it is small enough) for $c\in (0.5, 3)$ such that $\mathcal{A}^\alpha_n(\scrd{f}{ctm1},c)$ is a good contractive mapping. For $(\alpha, n)=(0.10, 30)$ and $(\alpha,n) = (0.05, 30)$, there are two intersection points for the line $y=c$ and the curve $y= \mathcal{A}^\alpha_n(\scrd{f}{ctm1},c)$, and the solutions $c^\alpha_n < 0.5$ are discarded. For $(\alpha,n) = (0.02, 30)$ and $(\alpha, n) = (0.01, 30)$, there is just one intersection point for the line $y=c$ and the curve  $y= \mathcal{A}^\alpha_n(\scrd{f}{ctm1},c)$, and the feasible solution $c^\alpha_n$ is larger than $1.5$. It is obvious that for the fixed $n$, the smaller the $\alpha$ is, the larger the $c^\alpha_n$ is.   

\subsubsection{Newton's Iterative Method}
The Newton's iterative formula for Kuiper's $V_n$-test can be written by
\begin{equation}
c_{i+1} = \mathcal{B}^\alpha_n(\scrd{f}{nlm1}, c_i)
\end{equation}
where
\begin{equation}
\mathcal{B}^\alpha_n(\scrd{f}{nlm1}, c) = c - \frac{\scrd{f}{nlm1}(c, \alpha, n)}{\scrd{f}{nlm1}(c, \alpha, n)},
\end{equation}
is the Newton's updating function. 
The mapping $\mathcal{B}^\alpha_n(\scrd{f}{nlm1}, c)$ is a contractive mapping for appropriate domain of $c$, say $c\in (1, 2)$.
The initial value for the Newton's iterative method could be set by $\scrd{c}{guess} = 1.2$. 

\begin{figure}[htbp]
\centering
\subfigure[$\alpha=0.10, n = 30, c^\alpha_n = 1.5503$]{
\includegraphics[width=0.38\textwidth]{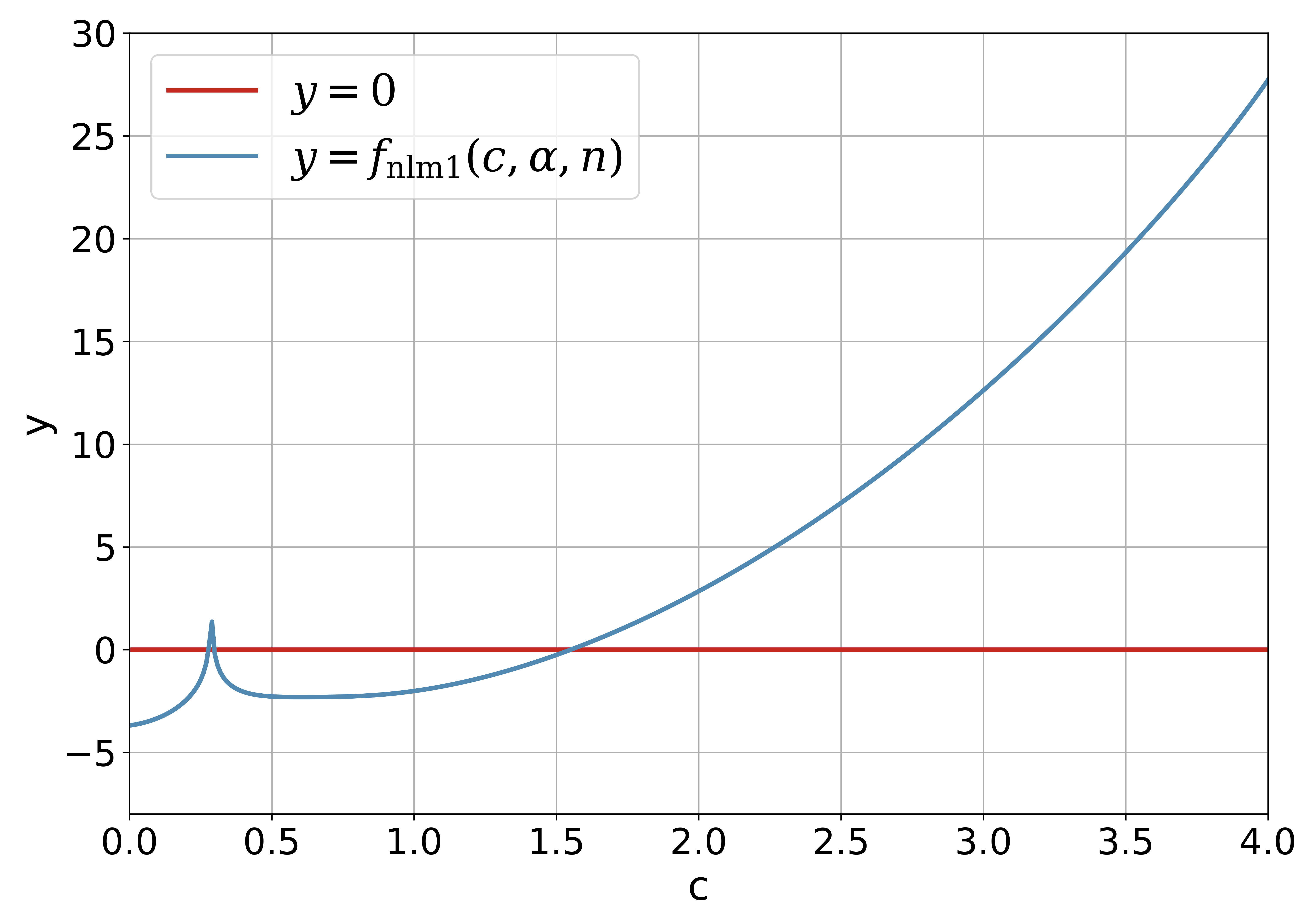} 
} 
\subfigure[$\alpha=0.05, n = 30, c^\alpha_n = 1.6758$]{
\includegraphics[width=0.38\textwidth]{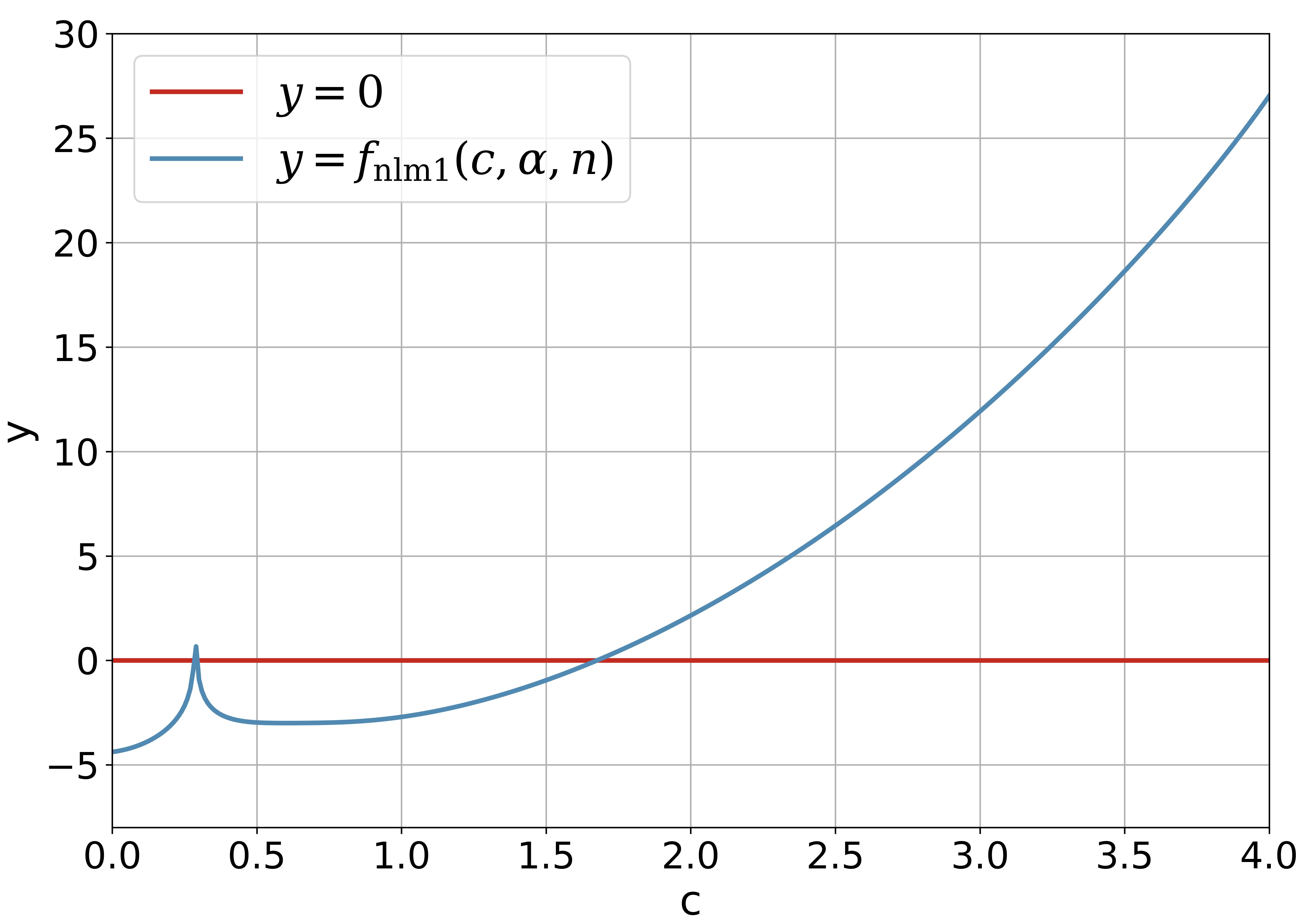} 
}
\subfigure[$\alpha=0.02, n = 30, c^\alpha_n = 1.8235$]{
\includegraphics[width=0.38\textwidth]{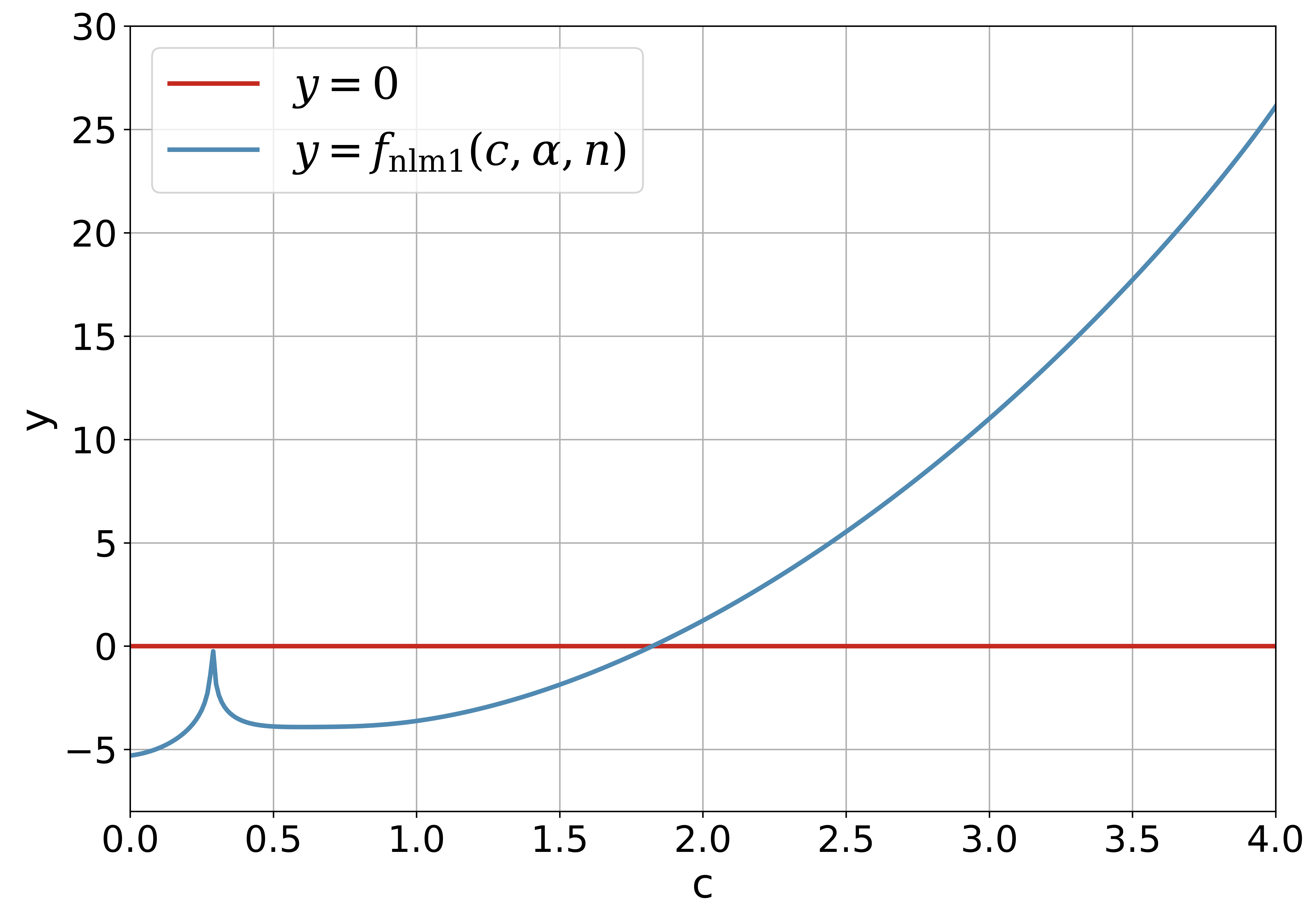} 
}
\subfigure[$\alpha=0.01, n = 30, c^\alpha_n = 1.9252$]{
\includegraphics[width=0.38\textwidth]{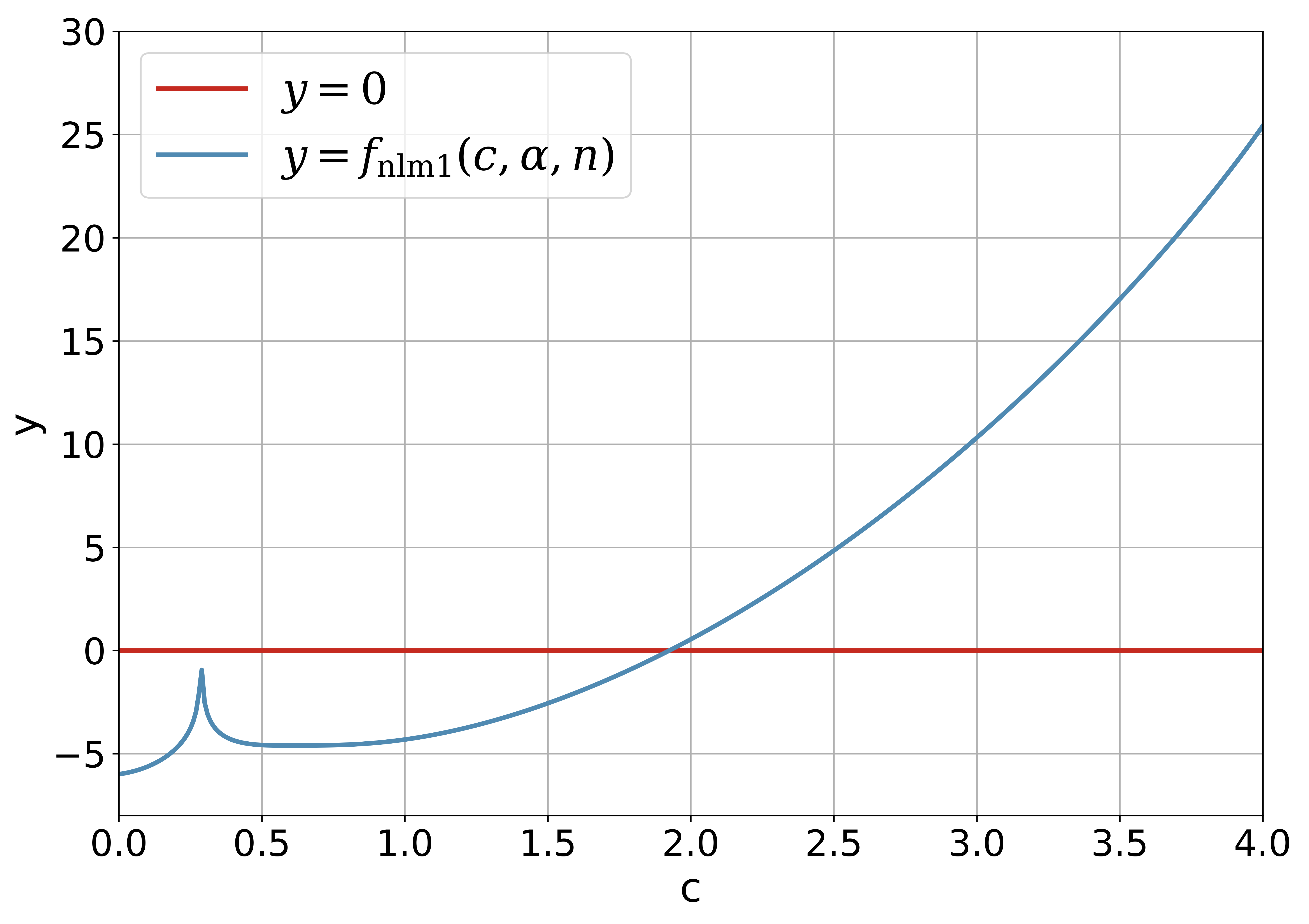} 
}
\caption{Curve and root of $\scrd{f}{nlm1}(c, \alpha, n)$ for $\alpha \in \set{0.10, 0.05, 0.02, 0.01}$ and $n = 30$}
\label{fig-Ganc-curve}
\end{figure}

\Fig \ref{fig-Ganc-curve} shows the curve of $\scrd{f}{nlm1}(c, \alpha, n)$ for $c\in(0, 3)$ and fixed $\alpha$ and $n$. It should be noted that  the necessary condition $c > 1/2$ automatically rejects the possible root which lies in the open interval $(0, 0.5)$. Although the diagram in \Fig \ref{fig-Ganc-curve}  is different its counterpart in \Fig \ref{fig-Aalphan-c}, the solution $c^\alpha_n$ obtained with the Newton's iterative method via $\scrd{f}{nlm1}(c,\alpha,n)$ is the same as that obtained by the direct iterative method via $\scrd{f}{ctm1}(c,\alpha,n)$.

\subsubsection{Kuiper's Pair for Kuiper's $V_n$-test}

With the help of the two contractive mappings  $\mathcal{A}^\alpha_n(\scrd{f}{ctm1}, \cdot)$ and $\mathcal{B}^\alpha_n(\scrd{f}{nlm1}, \cdot)$, the fixed-point $
c^\alpha_n$ for the equation \eqref{eq-utp-alpha-c} can be solved with the direct iterative method or Newton's iterative method, which is  the critical value such that
$\alpha = \Pr\set{K_n > c^\alpha_n}$. 
The relation of the critical value $c^\alpha_n$ and upper tail probability $v^\alpha_n$ can be expressed by
\begin{equation} \label{eq-c2v}
v^\alpha_n = \frac{c^\alpha_n}{\sqrt{n}}
\end{equation}
since we have 
\begin{equation}
\alpha = \Pr\set{\sqrt{n}\cdot V_n > c^\alpha_n} = \Pr\set{ V_n > c^\alpha_n/\sqrt{n}} = \Pr\set{V_n > v^\alpha_n}
\end{equation}
by the definitions of critical value and upper tail probability.  Once the critical value is obtained, the computation of upper tail quantile can be solved according to \eqref{eq-c2v}. This completes the computation of the Kuiper's pair $\mpair{c^\alpha_n}{v^\alpha_n}$ with the fixed-point method.

\subsection{Computation of the Pairs for Kuiper's $V_{n,n}$-Test}
\label{subsec-comp-method-Kuiper-Vnn}

For the  Kuiper's $V_{n,n}$-test, there is also a formula for the upper tail probability \cite{Kuiper1960TestsCR,Kemperman1959}
\begin{equation} \label{eq-alpha-c-Vnn}
1-\alpha = \Pr\set{\sqrt{n}\cdot V_{n,n}\le c} = 1- \sum^\infty_{j=1}2(2j^2c^2-1)\me^{-j^2c^2}
  + \frac{1}{6n}\left[1 + \sum^\infty_{j=1}j^2c^2(2j^2c^2-7)\me^{-j^2c^2} \right] + \BigO{\frac{1}{n^2}}.
\end{equation}
We can approximate the infinite series with the first and second term as done for the $V_n$-test, which implies that
\begin{equation}
\begin{aligned}
\alpha &\approx 2(2c^2-1)\me^{-c^2} + 2(8c^2-1)\me^{-4c^2}
      - \frac{1}{6n}\left[ 1 + c^2(2c^2-7)\me^{-c^2} + 4c^2(8c^2-7)\me^{-4c^2} \right]\\
       &= -\frac{1}{6n} + \left[2(2c^2-1) - \frac{c^2(2c^2-7)}{6n}\right]\me^{-c^2} 
       + \left[ 2(8c^2-1) -\frac{2c^2(8c^2-7)}{3n} \right]\me^{-4c^2}.
\end{aligned}
\end{equation}
Let
\begin{equation}
U_1(c,n) = 2(2c^2-1) - \frac{c^2(2c^2-7)}{6n}  -\frac{\me^{c^2}}{6n}, \quad 
U_2(c,n) = 2(8c^2-1) -\frac{2c^2(8c^2-7)}{3n},
\end{equation}
then we can deduce that
\begin{equation}
\boxed{c^2 + \ln \alpha = \ln\left[U_1(c,n) + U_2(c, n)\me^{-3c^2}\right]}.
\end{equation}
Let
\begin{equation}
\scrd{f}{nlm2}(c, \alpha, n) = c^2 + \ln \alpha -  \ln\left[U_1(c,n) + U_2(c, n)\me^{-3c^2}\right]
\end{equation}
where the digit $2$ in the subscript "nlm2" denotes the double $n$ in $V_{n,n}$ 
and 
\begin{equation}
\mathcal{A}^\alpha_{n,n}(\scrd{f}{ctm2}, c) = \scrd{f}{ctm2}(c, \alpha, n) = \sqrt{\ln\left[U_1(c,n) + U_2(c, n)\me^{-3c^2}\right] - \ln \alpha}
\end{equation}
We also can construct two iterative formulas for solving the nonlinear equation
\begin{equation}
\scrd{f}{nlm2}(c, \alpha, n) = 0
\end{equation}
to solve the $c^\alpha_{n,n}$ as follows:
\begin{itemize}
\item Newton's iterative method:
\begin{equation}
c_{i+1} = \mathcal{B}^\alpha_{n,n}(\scrd{f}{nlm2}, c_i) = c_i - \frac{\scrd{f}{nlm2}(c_i, \alpha, n)}{\scrd{f}{nlm2}'(c_i, \alpha, n)}
\end{equation}
\item direct iterative method:
\begin{equation}
c_{i+1} = \mathcal{A}^\alpha_{n,n}(\scrd{f}{ctm2}, c_i)  = \scrd{f}{ctm2}(c_i, \alpha, n) = \sqrt{\ln\left[U_1(c_i,n) + U_2(c_i,n)\me^{-3c_i^2}\right] - \ln \alpha}
\end{equation}
\end{itemize}
Just like the process of solving $c^\alpha_n$, the $c^\alpha_{n,n}$ defined by
\begin{equation}
\alpha = \Pr\set{\sqrt{n}\cdot V_{n,n} > c^\alpha_{n,n}}
\end{equation}
is the limit of the sequence $c_0, c_1, c_2, \cdots $ generated by $\mathcal{A}^\alpha_{n,n}(\scrd{f}{ctm2}, c)$ and $\mathcal{B}^\alpha_{n,n}(\scrd{f}{nlm2}, c)$. 
The initial value for solving $c^\alpha_{n,n}$ can be observed from the figure of $y=\scrd{f}{nlm2}(c, \alpha, n)$ with the variable $c$ intuitively. It should be noted that there is no modified version of the $V_{n,n}$-test due to the error bound for the approximation in \eqref{eq-alpha-c-Vnn} is $\BigO{n^{-2}}$ instead of $\BigO{n^{-1}}$.

\begin{figure}[htbp] 
\centering
\subfigure[$\alpha=0.10, n = 30, c^\alpha_n = 2.2740$]{
\includegraphics[width=0.38\textwidth]{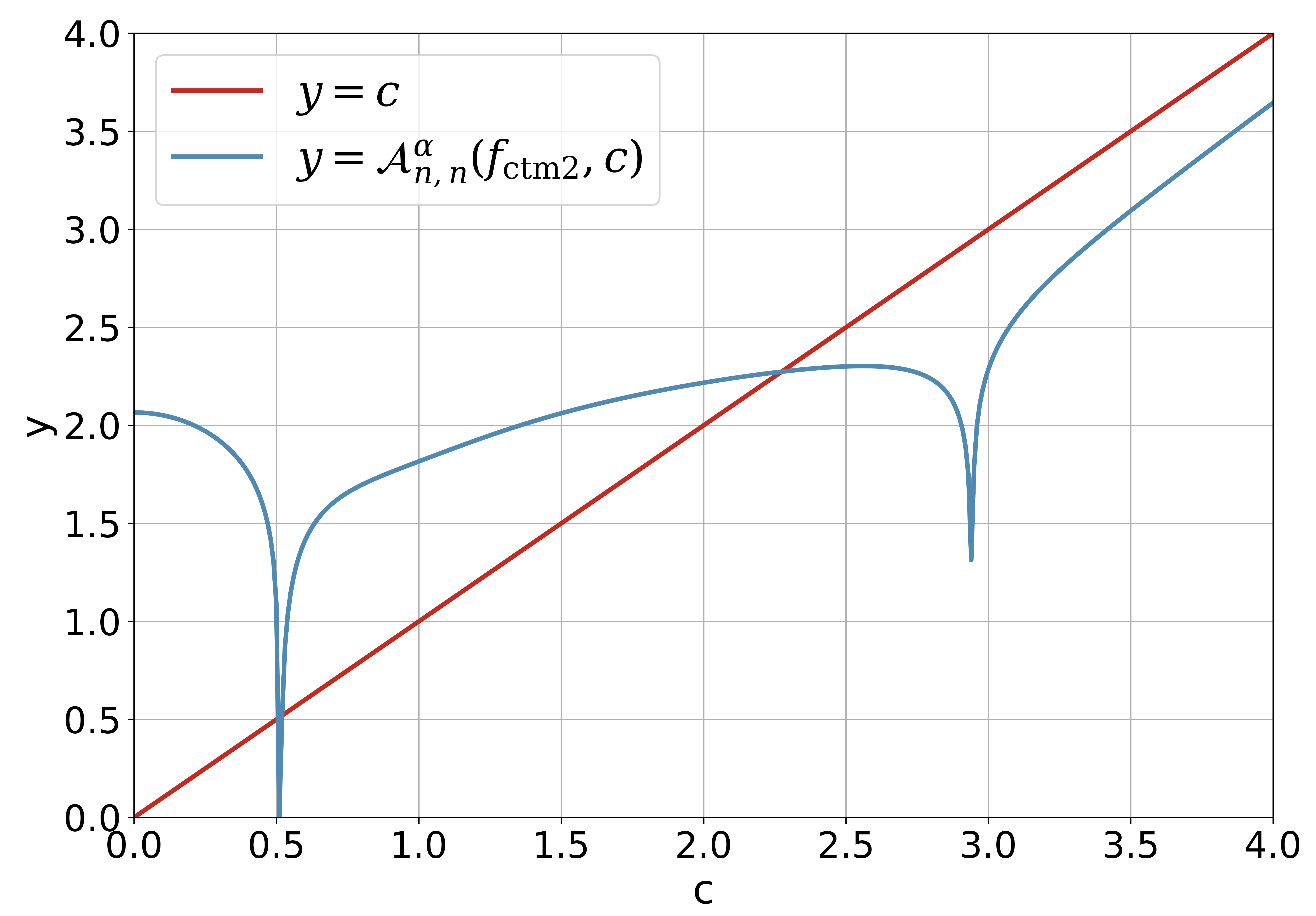} 
} 
\subfigure[$\alpha=0.05, n = 30, c^\alpha_n = 2.4430$]{
\includegraphics[width=0.38\textwidth]{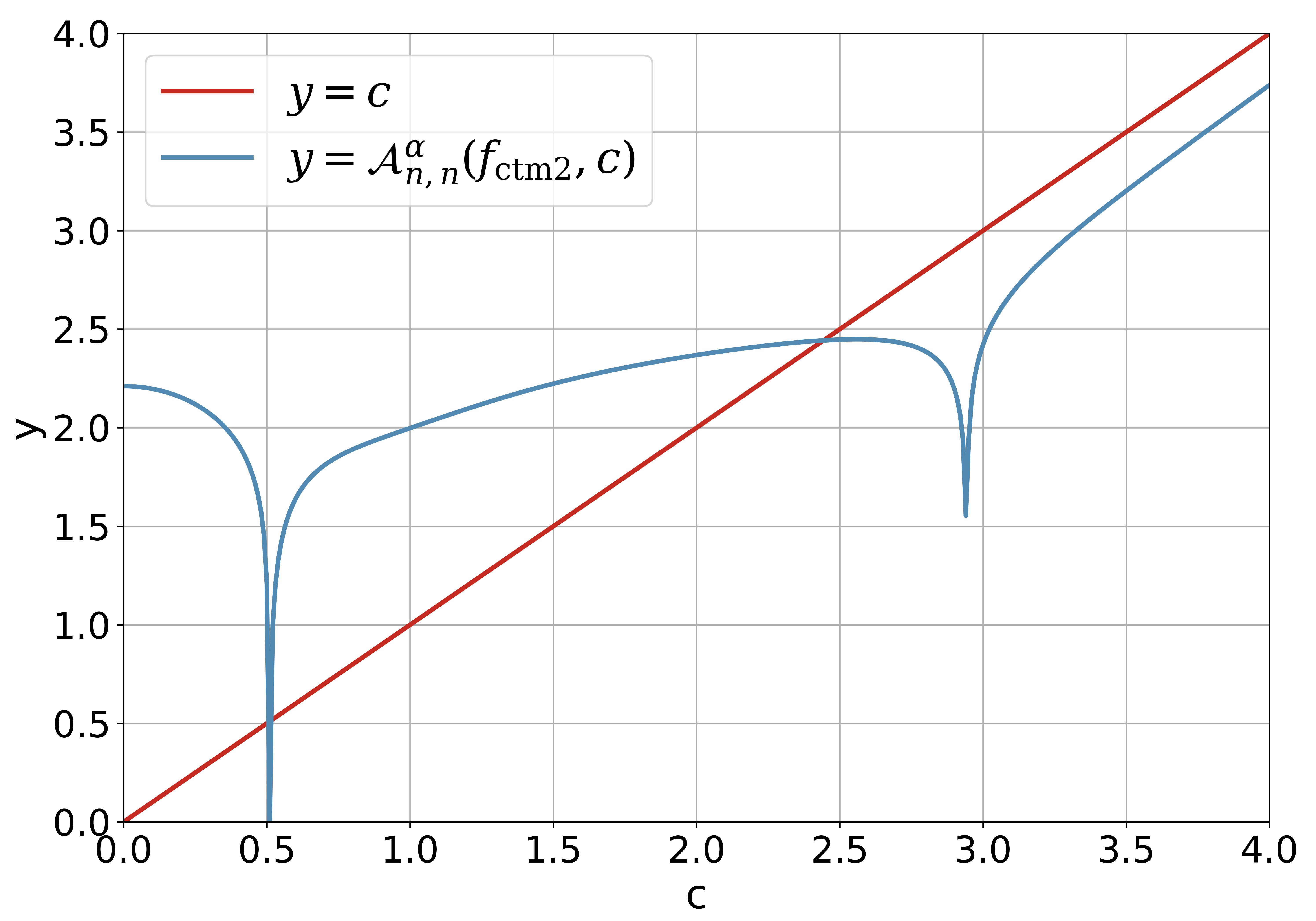} 
}
\subfigure[$\alpha=0.02, n = 30, c^\alpha_n = 2.6266$]{
\includegraphics[width=0.38\textwidth]{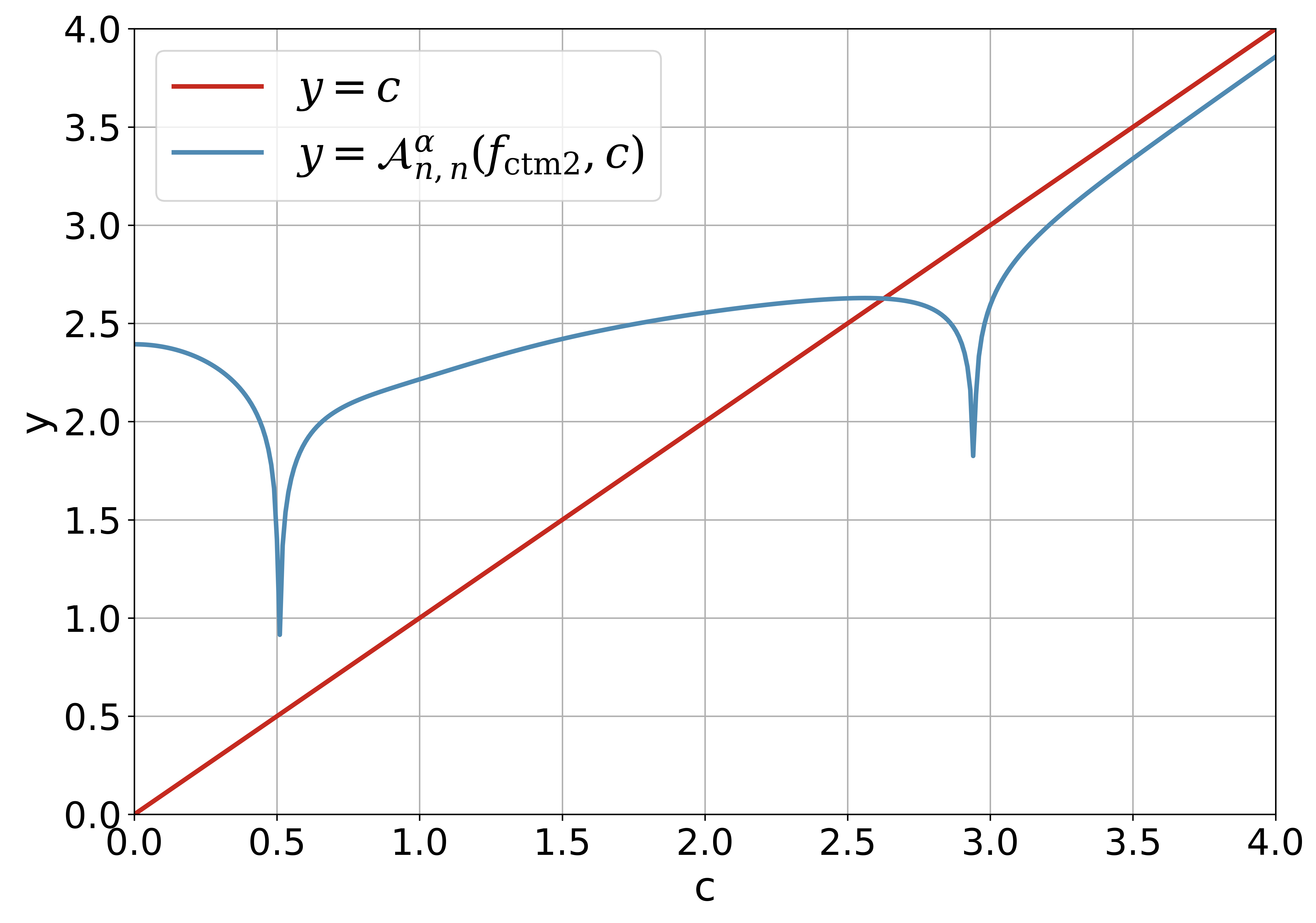} 
}
\subfigure[$\alpha=0.01, n = 30, c^\alpha_n = 2.7351$]{
\includegraphics[width=0.38\textwidth]{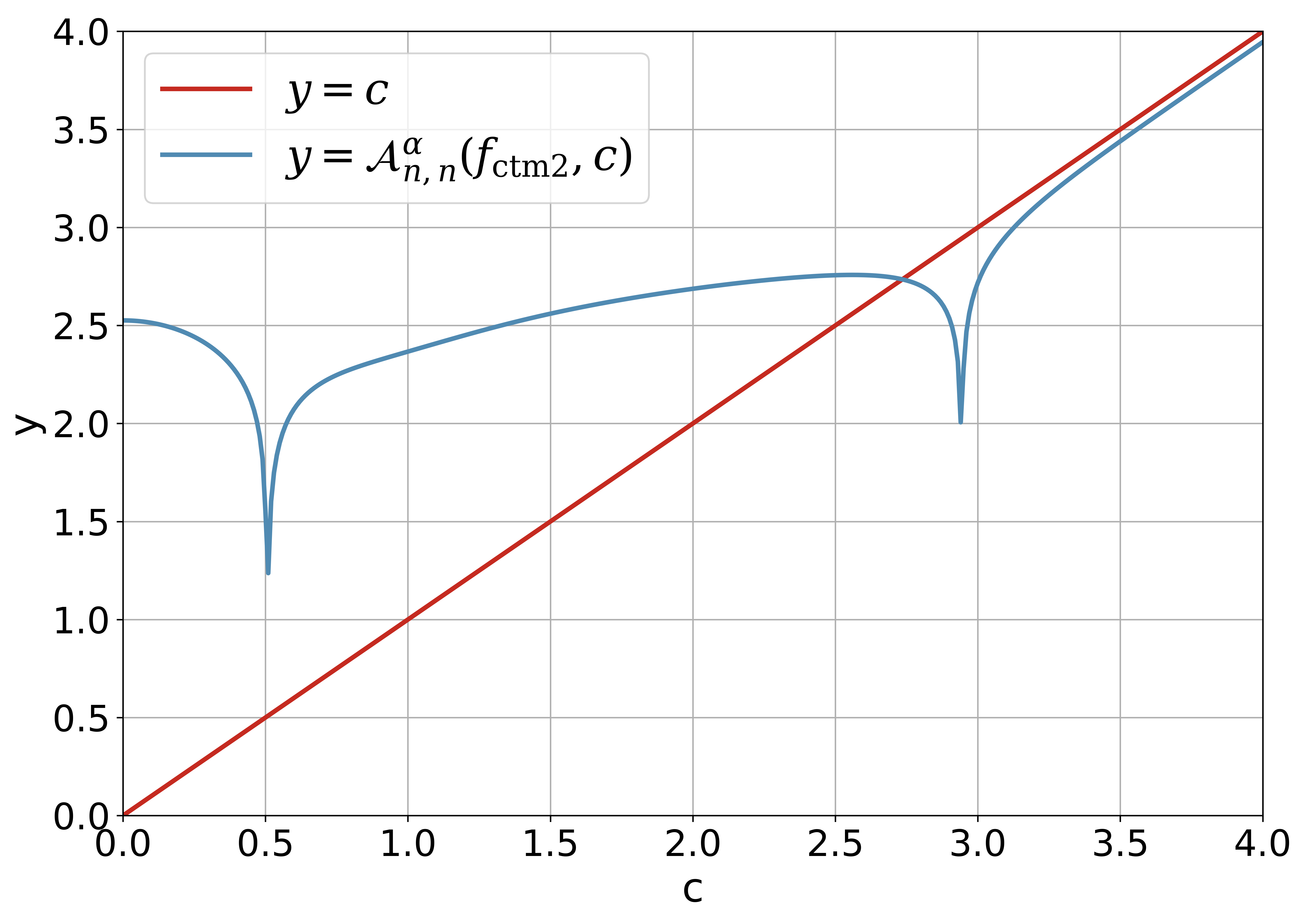} 
}
\caption{Intersection  of  $y= c$ and $ y = \mathcal{A}^\alpha_{n,n}(\scrd{f}{ctm2},\alpha,n) = \scrd{f}{ctm2}(c, \alpha, n)$ for $\alpha\in \set{0.10, 0.05, 0.02, 0.01}$ and $n= 30$.}\label{fig-vnn-UTP-fctm2}
\end{figure}

\begin{figure}[htbp] 
\centering
\subfigure[$\alpha=0.10, n = 30, c^\alpha_n = 2.2740$]{
\includegraphics[width=0.38\textwidth]{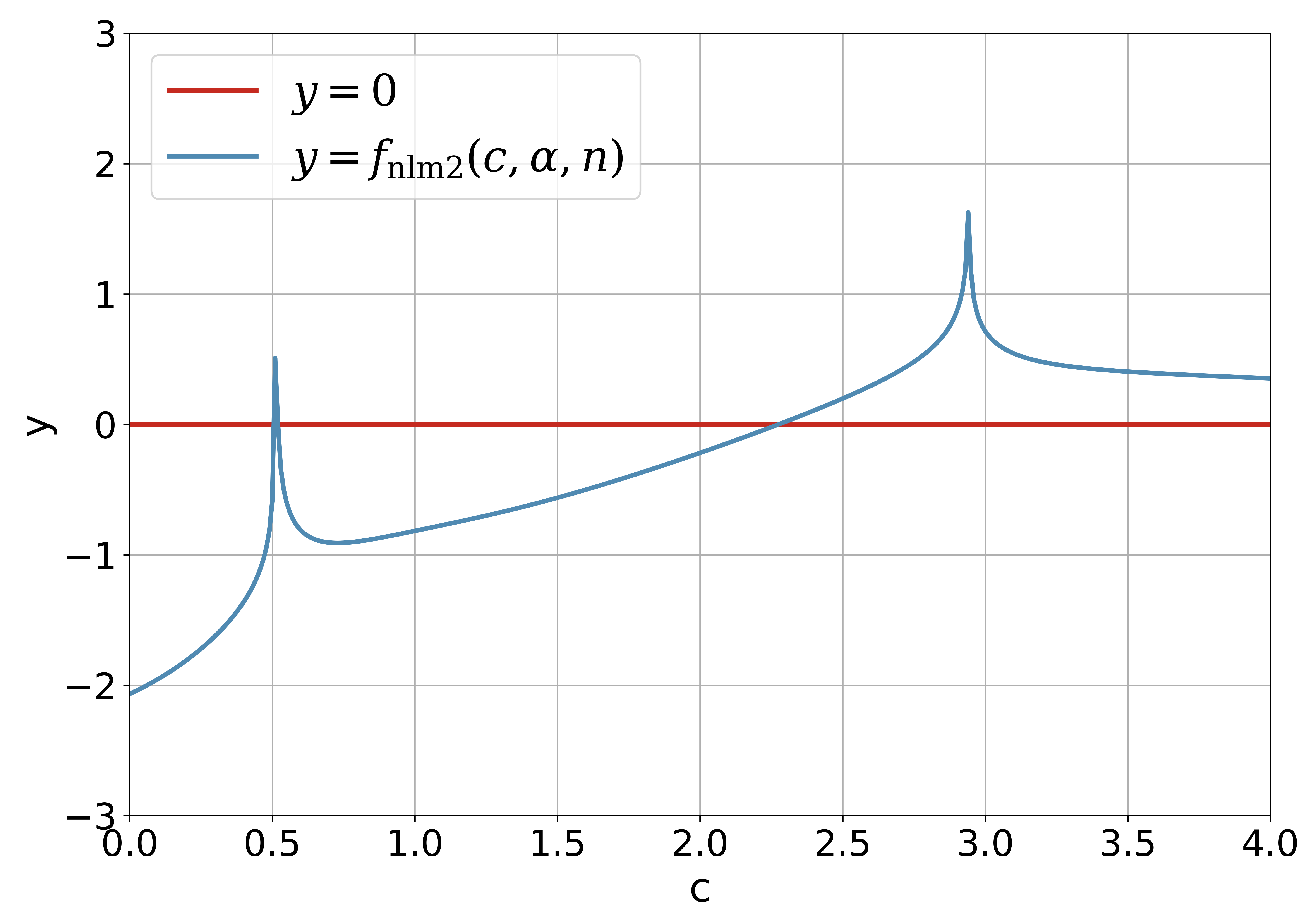} 
} 
\subfigure[$\alpha=0.05, n = 30, c^\alpha_n = 2.4430$]{
\includegraphics[width=0.38\textwidth]{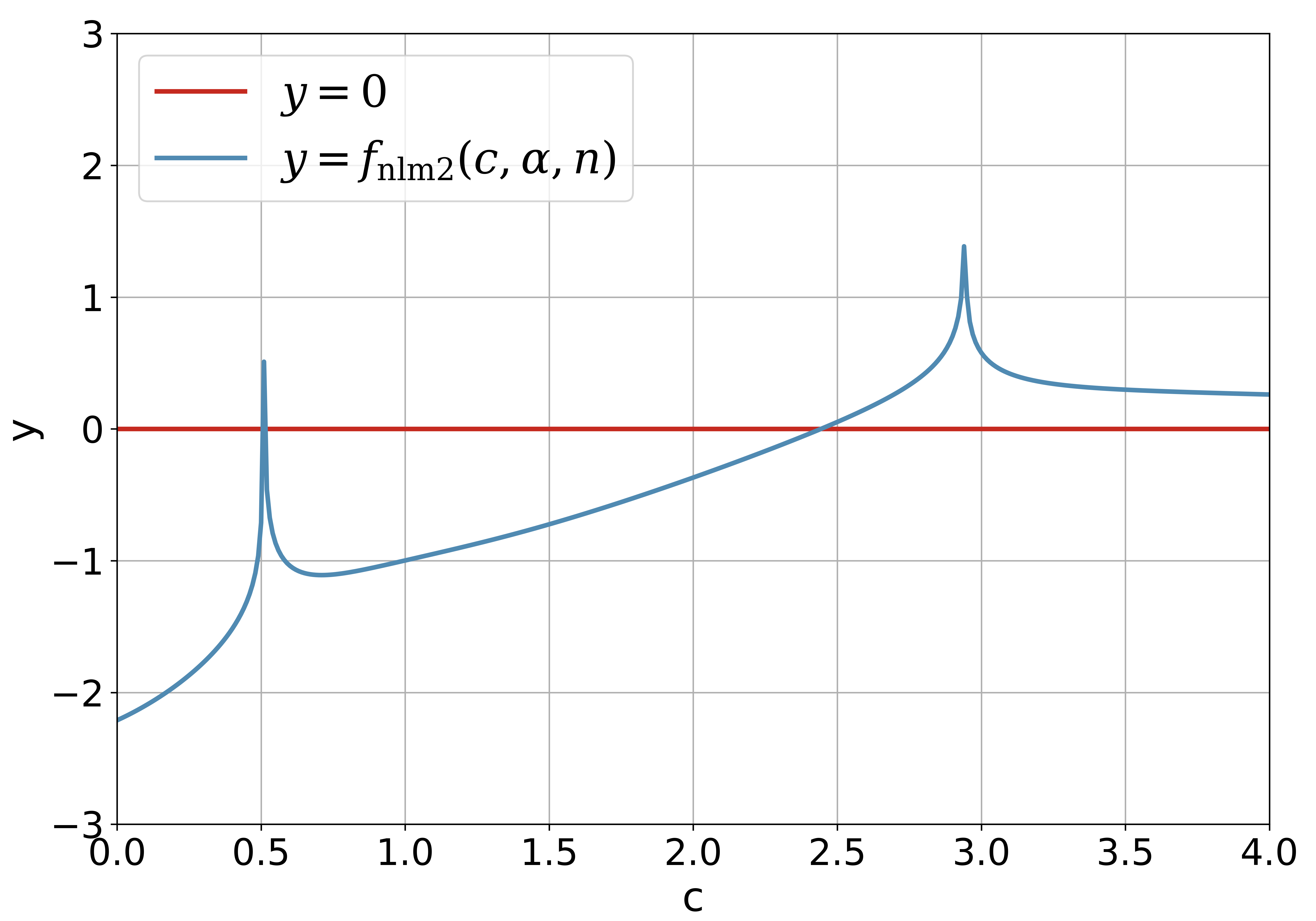} 
}
\subfigure[$\alpha=0.02, n = 30, c^\alpha_n = 2.6266$]{
\includegraphics[width=0.38\textwidth]{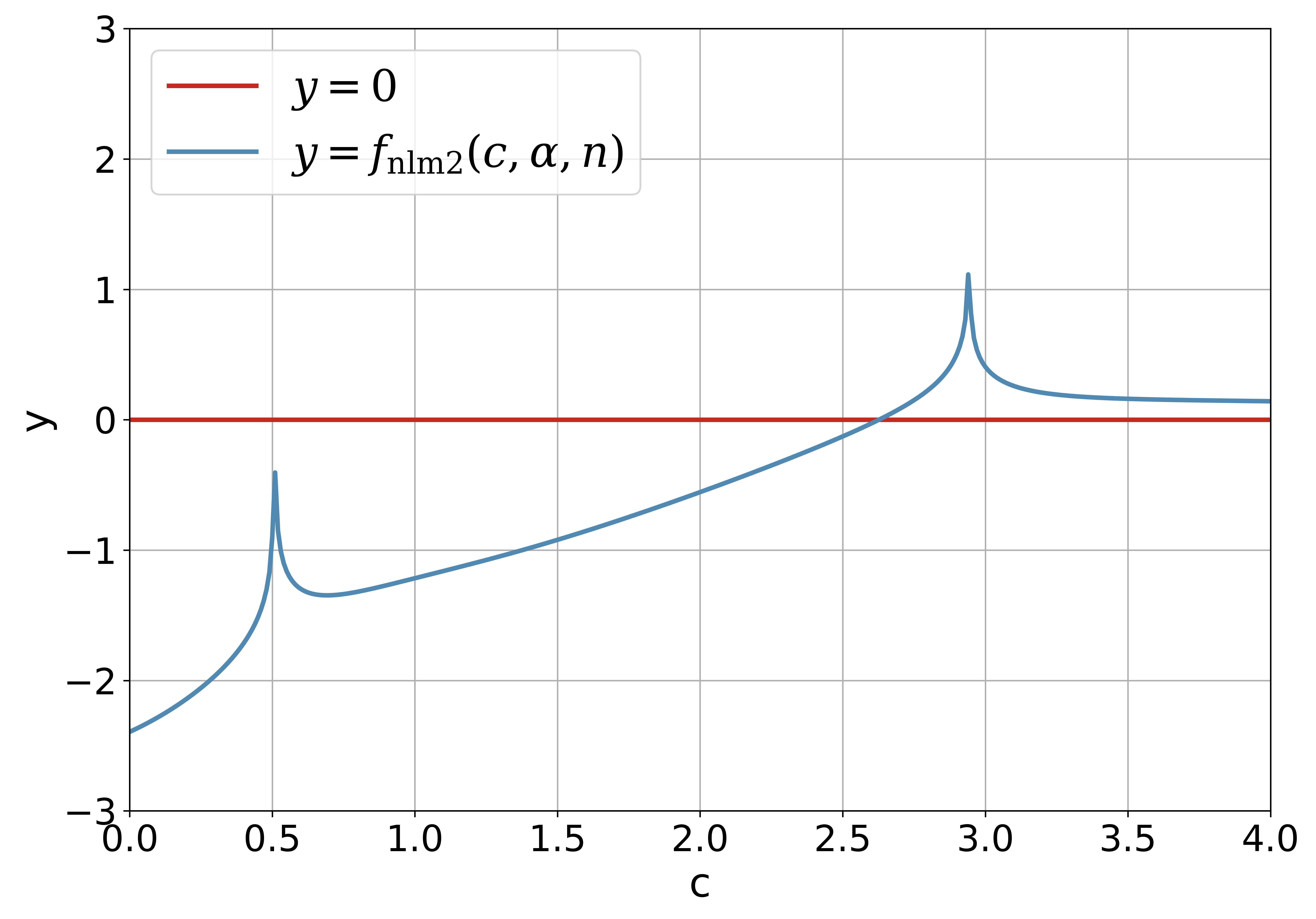} 
}
\subfigure[$\alpha=0.01, n = 30, c^\alpha_n = 2.7351$]{
\includegraphics[width=0.38\textwidth]{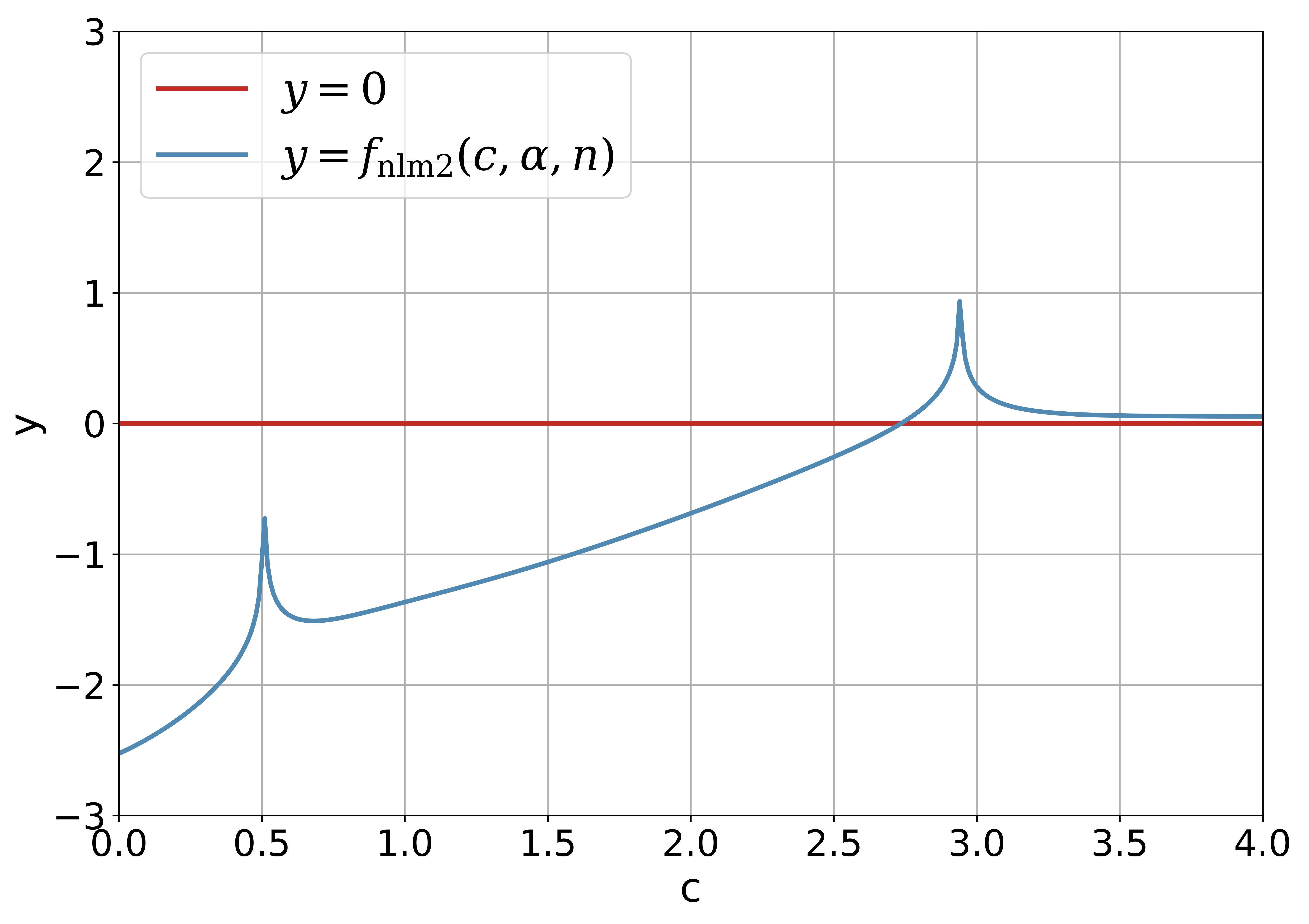} 
}
\caption{Curve and root of $\scrd{f}{nlm2}(c, \alpha, n) = 0$ for $\alpha\in \set{0.10, 0.05, 0.02, 0.01}$ and $n= 30$.}\label{fig-vnn-UTP-fnlm2}
\end{figure}

As an illustration, \Fig \ref{fig-vnn-UTP-fctm2} shows the critical value $c^\alpha_{n,n}$ of the statistic $V_{n,n}$ with the direct iterative method via $\scrd{f}{ctm2}(c,\alpha,n)$, in which $\alpha \in \set{0.10, 0.05, 0.02, 0.01}$ and $n=30$. Similarly,  \Fig \ref{fig-vnn-UTP-fnlm2} shows the counterpart of  $c^\alpha_{n,n}$ with the Newton's iterative method via $\scrd{f}{nlm2}(c,\alpha,n)$. It should be noted that when the $\alpha$ decreases, the critical value $c^\alpha_{n,n}$ increases.

\section{Fixed-Point Algorithms for Solving the Pair of Kuiper's Test}
\label{sec-algorithms}

\subsection{Auxiliary Procedures}

In order to reduce the structure complexity of our key algorithms for solving the Kuiper's pair $\mpair{c^\alpha_n}{v^\alpha_n}$, it is wise to introduce some auxiliary procedures. The procedures \ProcName{FunA1cn}  and \ProcName{FunA2cn} in  \Algor \ref{alg-calc-A1nc} and
\Algor \ref{alg-calc-A2nc}  are designed for computing $A_1(c,n)$ and $A_2(c,n)$ respectively.

\begin{breakablealgorithm}
\caption{Calculating the value of $A_1(c, n)$ arising in the updating function in $V_n$-test}
\label{alg-calc-A1nc}
\begin{algorithmic}[1]
\Require  Positive critical value $c$, positive integer $n$ 
\Ensure  The value of $A_1(c, n)$.
\Function{FunA1cn}{$c, n$}
\State $ A_1 \gets -2 + 8c/\sqrt{n} + 8c^2 -32c^3/(3\sqrt{n})$;
\State \Return $A_1$;
\EndFunction
\end{algorithmic}
\end{breakablealgorithm}

\begin{breakablealgorithm}
\caption{Calculating the value of $A_2(c, n)$ arising in the  $V_n$-test}
\label{alg-calc-A2nc}
\begin{algorithmic}[1]
\Require  Positive critical value $c$,  positive integer $n$ 
\Ensure  The value of $A_2(c, n)$.
\Function{FunA2cn}{$c, n$}
\State $ A_2 \gets -2 + 32c/\sqrt{n} + 32c^2 -512c^3/(3\sqrt{n})$;
\State \Return $A_2$;
\EndFunction
\end{algorithmic}
\end{breakablealgorithm}

The procedures \ProcName{FunFnlm1}  and \ProcName{FunFnlm2} in  \Algor \ref{alg-f-nlm-1} and
\Algor \ref{alg-f-nlm-2}  are designed for computing the non-linear functions $\scrd{f}{nlm1}(c, \alpha, n)$ for the $V_n$-test and $\scrd{f}{nlm2}(c, \alpha, n)$ for the $V_{n,n}$-test respectively.

\begin{breakablealgorithm}
\caption{Calculating the value of $\scrd{f}{nlm1}(c, \alpha, n)$ for the updating function in $V_n$-test}
\label{alg-f-nlm-1}
\begin{algorithmic}[1]
\Require Positive critical value $c$, upper tail probability $\alpha$,  integer $n$ 
\Ensure  The value of $\scrd{f}{nlm1}(c, \alpha, n)$.
\Function{FunFnlm1}{$c, \alpha, n$}
\State $ A_1 \gets \ProcName{FunA1cn}(c,n)$;
\State $ A_2 \gets \ProcName{FunA2cn}(c,n)$;
\State $ u \gets 2c^2 + \ln \alpha - \ln\left(A_1 + A_2\cdot \me^{-6c^2}\right)$;
\State \Return $u$;
\EndFunction
\end{algorithmic}
\end{breakablealgorithm}

\begin{breakablealgorithm}
\caption{Calculating the value of $\scrd{f}{nlm2}(c, \alpha, n)$ for the updating function in $V_{n,n}$-test}
\label{alg-f-nlm-2}
\begin{algorithmic}[1]
\Require Positive critical value $c$, upper tail probability $\alpha$,  integer $n$ 
\Ensure  The value of $\scrd{f}{nlm2}(c, \alpha, n)$.
\Function{FunFnlm2}{$c, \alpha, n$}
\State $x \gets c^2$; 
\State $ U_1 \gets 2(2x-1)-x(2x-7)/(6n) - \me^{x}/(6n)$;
\State $ U_2 \gets 2(8x-1)-2x(8x-7)/(3n)$;
\State $ u \gets x + \ln \alpha - \ln\left(U_1 + U_2\cdot \me^{-3x}\right)$;
\State \Return $u$;
\EndFunction
\end{algorithmic}
\end{breakablealgorithm}

The procedures \ProcName{FunFctm1}  and \ProcName{FunFctm2} in  \Algor \ref{alg-f-ctm-1} and
\Algor \ref{alg-f-ctm-2}  are designed for computing the non-linear functions $\scrd{f}{ctm1}(c, \alpha, n)$ for the $V_n$-test and $\scrd{f}{nct2}(c, \alpha, n)$ for the $V_{n,n}$-test respectively.

\begin{breakablealgorithm}
\caption{Calculating the $\scrd{f}{ctm1}(c,\alpha,n)$ for the updating operator $\mathcal{A}^\alpha_n(f, c)$ in $V_n$-test }
\label{alg-f-ctm-1}
\begin{algorithmic}[1]
\Require  Positive critical value $c$, upper tail probability $\alpha$, integer $n$
\Ensure  The value of the direct updating function $\mathcal{A}^\alpha_n(c)$.
\Function{FunFctm1}{$c, \alpha, n$}
\State $ A \gets \ProcName{FunA1cn}(c,n)$;
\State $ B \gets \ProcName{FunA2cn}(c,n)$;
\State $ y \gets\sqrt{\left(\ln(A + B \cdot \me^{-6c^2}) -\ln \alpha \right)/2}$; 
\State \Return $y$;
\EndFunction
\end{algorithmic}
\end{breakablealgorithm}

\begin{breakablealgorithm}
\caption{Calculating the $\scrd{f}{ctm2}(c,\alpha,n)$ for the updating operator $\mathcal{A}^\alpha_{n,n}(f,c)$ in $V_{n,n}$-test }
\label{alg-f-ctm-2}
\begin{algorithmic}[1]
\Require Positive critical value $c$, pper tail probability $\alpha$, integer $n$
\Ensure  The value of the $\scrd{f}{ctm2}(c,\alpha,n)$.
\Function{FunFctm2}{$c, \alpha, n$}
\State $x \gets c^2$; 
\State $ U_1 \gets 2(2x-1)-x(2x-7)/(6n) - \me^{x}/(6n)$;
\State $ U_2 \gets 2(8x-1)-2x(8x-7)/(3n)$;
\State $ y\gets \sqrt{\ln(U_1 + U_2 \cdot \me^{-3x}) - \ln \alpha}$;
\State \Return $y$;
\EndFunction
\end{algorithmic}
\end{breakablealgorithm}

\subsection{Updating Mapping }

The updating mapping is essential for the iterative scheme for solving the fixed-point. The high order procedure \ProcName{UpdateMethodDirect}
in \Algor \ref{alg-Direct-updator} is designed for computing the contractive mapping $\mathcal{A}^\alpha_n(f, c)$  for the $V_n$-test and $\mathcal{A}^\alpha_{n,n}(f, c)$ for the $V_{n,n}$-test with a unified interface for the direct iterative method. 
\begin{breakablealgorithm}
\caption{Calculating the Contractive Mapping $\mathcal{A}^\alpha_n(f, c)$ and $\mathcal{A}^\alpha_{n,n}(f, c)$  }
\label{alg-Direct-updator}
\begin{algorithmic}[1]
\Require  Contractive function $\scrd{f}{ctm}(c, \alpha, n)$, critical value $c$, upper tail probability $\alpha$, integer $n$
\Ensure  The value of the direct updating function $\mathcal{A}^\alpha_n(\scrd{f}{ctm1}, c)$ or $\mathcal{A}^\alpha_{n,n}(\scrd{f}{ctm2}, c)$. 
   \begin{itemize}
   \item if $\scrd{f}{ctm} = \scrd{f}{ctm1}$, then we get $\mathcal{A}^\alpha_n(\scrd{f}{ctm1},c)$ in $V_n$-test and 
   \item if $\scrd{f}{ctm} = \scrd{f}{ctm2}$, then we get $\mathcal{A}^\alpha_{n,n}(\scrd{f}{ctm2},c)$ in $V_{n,n}$-test.
  \end{itemize}
\Function{UpdateMethodDirect}{$\scrd{f}{ctm}, c, \alpha, n$}
\State $ u \gets \scrd{f}{ctm}(c, \alpha, n)$; 
\State \Return $u$;
\EndFunction
\end{algorithmic}
\end{breakablealgorithm}

Similarly, 
The high order procedure \ProcName{UpdateMethodNewton}
in \Algor \ref{alg-Newton-updator} is designed for computing the contractive mapping $\mathcal{B}^\alpha_n(f, c)$  for the $V_n$-test and $\mathcal{B}^\alpha_{n,n}(f, c)$ for the $V_{n,n}$-test with a unified interface for the Newton iterative method.

\begin{breakablealgorithm}
\caption{Calculating the Newton's updating functions $\mathcal{B}^\alpha_n(\scrd{f}{nlm}, c)$ and $\mathcal{B}^\alpha_{n,n}(\scrd{f}{nlm}, c)$  }
\label{alg-Newton-updator}
\begin{algorithmic}[1]
\Require Nonlinear function $\scrd{f}{nlm}(\alpha, n, c)$, upper tail probability $\alpha$, integer $n$, critical value $c$
\Ensure  The value of the Newton's updating function $\mathcal{B}^\alpha_n(\scrd{f}{nlm},c)$ or $\mathcal{B}^\alpha_{n,n}(\scrd{f}{nlm}, c)$. 
   \begin{itemize}
   \item if $\scrd{f}{nlm}= \scrd{f}{nlm1}$, then we get $\mathcal{B}^\alpha_n(\scrd{f}{nlm1}, c)$ in $V_n$-test and 
   \item if $\scrd{f}{nlm}= \scrd{f}{nlm2}$, then we get $\mathcal{B}^\alpha_{n,n}(\scrd{f}{nlm2}, c)$ in $V_{n,n}$-test.
  \end{itemize}
\Function{UpdateMethodNewton}{$\scrd{f}{nlm}, \alpha, n, c$}
\State $ h \gets 10^{-5}$; 
\State $\cpvar{slope} \gets  \left(\scrd{f}{nlm}(c+h, \alpha, n) - \scrd{f}{nlm}(c, \alpha, n)\right)/h$; // Calculate the derivative  $\scrd{f}{nlm}'(c, \alpha, n)$
\State $\scrd{c}{new} \gets c - \scrd{f}{nlm}(c, \alpha, n)/\cpvar{slope}$;
\State \Return $\scrd{c}{new}$;
\EndFunction
\end{algorithmic}
\end{breakablealgorithm}

\subsection{Algorithms for Solving the Kuiper Pair for $V_n$-test and $V_{n,n}$-test }

It is significant for us to calculate the original Kuiper pair $\mpair{c^\alpha_\star}{v^\alpha_\star}$ for Kuiper's $V_\star$-test where $V_\star \in \set{V_n, V_{n,n}}$. The procedure \ProcName{KuiperPairSolver} 
listed in \Algor \ref{alg-KuiperPairSolver} provides a unified framework for solving the Kuiper pair $\mpair{c^\alpha_\star}{v^\alpha_\star}$ with the Direct or Newton's iterative method.

\begin{breakablealgorithm}
\caption{Fixed-point iterative algorithm for solving the Kuiper's pair $\mpair{c^\alpha_n}{v^\alpha_n}$ in $V_n$-test or $\mpair{c^\alpha_{n,n}}{v^\alpha_{n,n}}$ in $V_{n,n}$-test with the while-do loop and direct/Newton iterative method}
\label{alg-KuiperPairSolver}
\begin{algorithmic}[1]
\Require  The number of samples $n$, upper tail probability $\alpha\in(0, 1)$, guess of the critical value $\scrd{c}{guess}$ with default value $\scrd{c}{guess}=2.45$, integer $\cpvar{method}\in \set{1,2}$ for the Direct/Newton iterative method with default value $\cpvar{method} = 2$ for the Newton's iterative method, integer $\cpvar{type}\in \set{1, 2}$ for the $V_n$-test and $V_{n,n}$-test with default value 1.  
\Ensure The Kuiper pair $\mpair{c^\alpha_\star}{v^\alpha_\star}$ such that $\alpha=\Pr\set{\sqrt{n}\cdot V_\star > c^\alpha_\star} = \Pr\set{V_\star \ge v^\alpha_\star}$ for $V_\star\in \set{V_n, V_{n,n}}$.
\Function{KuiperPairSolver}{$\scrd{c}{guess},\alpha, n,\cpvar{type},\cpvar{method}$}
\If{($\cpvar{method} == 1 \land \cpvar{type == 1}$)}
   \State $\ProcName{T} \gets \ProcName{UpdateMethodDirect}$; // using direct iterative method
   \State $f \gets \ProcName{FunFctm1}$; // Solve the fixed-point by $c = \mathcal{A}^\alpha_{n}(\scrd{f}{ctm1}, c)$ in $V_n$-test
\EndIf
\If{($\cpvar{method} == 1 \land \cpvar{type == 2}$)}
   \State $T \gets \ProcName{UpdateMethodDirect}$; // using direct iterative method
   \State $f\gets \ProcName{FunFctm2}$; // Solve the fixed-point by $c = \mathcal{A}^\alpha_{n,n}(\scrd{f}{ctm2},c)$ in $V_{n,n}$-test
\EndIf
\If{($\cpvar{method} == 2 \land \cpvar{type == 1}$)}
   \State $T \gets \ProcName{UpdateMethodNewton}$; // using Newton's iterative method
   \State $f\gets \ProcName{FunFnlm1}$; // solve the nonlinear eq. $\scrd{f}{nlm1}(c, \alpha, n) = 0$ in $V_n$-test
\EndIf
\If{($\cpvar{method} == 2 \land \cpvar{type == 2}$)}
   \State $T \gets \ProcName{UpdateMethodNewton}$; // using Newton's iterative method
   \State $f\gets \ProcName{FunFnlm2}$; // solve the nonlinear eq. $\scrd{f}{nlm2}(c, \alpha, n) = 0$ in $V_{n,n}$-test
\EndIf
\State $\epsilon \gets 10^{-5}$;
\State $d\gets \ProcName{Distance}$; // distance function
\State $c^\alpha_\star \gets \ProcName{FixedPointSolver}(T, f, d, \epsilon, \scrd{c}{guess}, \alpha, n)$;
\State $v^\alpha_\star \gets c^\alpha_\star/\sqrt{n}$;
\State \Return $\mpair{c^\alpha_\star}{v^\alpha_\star}$; // Kuiper pair for  the $V_\star$-test
\EndFunction
\end{algorithmic}
\end{breakablealgorithm}

\subsection{Algorithms for Solving the Upper/Lower Tail Quantile and Inverse of CDF}

For most applications, we may have no interest in the critical value $c^\alpha_n$ and our emphasis is put on the upper/lower tail quantile $v^\alpha_n = v_{1-\alpha}^n$ in $V_n$-test. Moreover, it is enough to choose the Newton's iterative method in order to find the upper/lower tail quantile. In \Algor \ref{alg-Kuiper-UTQ},
the procedure \ProcName{KuiperUTQ} is designed to solve the upper tail quantile. We remark that in the implementation of \ProcName{KuiperUTQ}, the procedure \ProcName{FixedPointSolver} is used instead of \ProcName{KuiperPairSolver}. 

\begin{breakablealgorithm}
\caption{Computing the upper tail quantile in Kuiper's $V_n$-test}
\label{alg-Kuiper-UTQ}
\begin{algorithmic}[1]
\Require Upper tail probability $\alpha\in(0, 1)$, capacity $n$ of the samples.  
\Ensure  Upper tail quantile $v^\alpha_n$ for the $V_n$ statistic such that $\alpha = \Pr\set{V_n > v}$. 
\Function{KuiperUTQ}{$\alpha, n$}
\If{$\alpha \ge  0.9999$}
\State \Return $0.0$;
\EndIf
\State $T \gets \ProcName{UpdateMethodNewton}$; // using Newton's iterative method
\State $f\gets \ProcName{FunFnlm1}$; \quad // solve the nonlinear eq. $\scrd{f}{nlm1}(c, \alpha, n) = 0$ in $V_n$-test
\State $\epsilon \gets 10^{-5}$; \quad // precision for the fixed-point
\State $d\gets \ProcName{Distance}$; \quad // distance function
\State $\scrd{c}{guess} \gets 2.45$; \quad // $\scrd{c}{guess}\in (1.1, 2.5)$ is OK.
\State $c^\alpha_n \gets \ProcName{FixedPointSolver}(T, f, d, \epsilon, \scrd{c}{guess}, \alpha, n)$;
\State $v^\alpha_n \gets  c^\alpha_n /\sqrt{n}$; 
\State \Return $v^\alpha_n$; 
\EndFunction
\end{algorithmic}
\end{breakablealgorithm}

The lower tail quantile may become more attractive for some applications.  The procedure 
\ProcName{KuiperLTQ} listed in \Algor \ref{alg-Kuiper-LTQ} is based on the procedures
\ProcName{KuiperPairSolver}. An alternative implementation 
of \ProcName{KuiperLTQ} can be done with the equivalence of $v_\alpha^n = v^{1-\alpha}_n$. In other words, we have $\ProcName{KuiperLTQ}(\alpha,n) \equiv \ProcName{KuiperUTQ}(1-\alpha, n)$ for any $\alpha \in [0, 1]$.

\begin{breakablealgorithm}
\caption{Compute the Lower Tail Quantile $v_\alpha^n$ in Kuiper's $V_n$-test}
\label{alg-Kuiper-LTQ}
\begin{algorithmic}[1]
\Require Lower tail probability $\alpha\in(0, 1)$, capacity $n$ of the samples.  
\Ensure  Lower tail quantile $v_\alpha^n$ in  Kuiper's $V_n$-test. 
\Function{KuiperLTQ}{$\alpha, n$}
\If{$\alpha \le  0.0001$}
\State \Return $0.0$;
\EndIf
\State $\scrd{c}{guess}\gets 2.45$;\quad // $\scrd{v}{guess}\in (1.1, 2.5)$ is OK
\State $\cpvar{type} \gets 1$; \quad // Kuiper's $V_n$-test
\State $\cpvar{method}\gets 2$; \quad // Newton's iterative method
\State $\mpair{c}{v}\gets \ProcName{KuiperPairSolver}(\scrd{c}{guess}, 1-\alpha, n, 
\cpvar{type}, \cpvar{method}) $;
\State \Return $v$; \quad // lower tail quantile, $v = v_{\alpha} = v^{1-\alpha}$ 
\EndFunction
\end{algorithmic}
\end{breakablealgorithm}

For the applications of goodness-of-fit test where Kuiper's $V_n$-test is involved, it is the
inverse of the CDF of $V_n$ that must be solved according to \eqref{eq-significance-level} or
\eqref{eq-inv-CDF}. It is easy to find that
\begin{equation}
v_x^n = \inv{F}_{V_n}(x), \quad \forall x\in [x,1]
\end{equation}
by the definitions of lower tail probability and CDF. Therefore, the inverse of CDF can be obtained by the 
calling the procedure \ProcName{KuiperLTQ} or \ProcName{KuiperUTQ} directly with proper argument. 
The procedure \ProcName{KuiperInvCDF} in \Algor \ref{alg-Kuiper-InvCDF} is designed for computing the  $\inv{F}_{V_n}(x)$ for any probability $x\in[0,1]$.

\begin{breakablealgorithm}
\caption{Compute the inverse CDF $\inv{F}_{V_n}(x)$ for Kuiper's $V_n$-test}
\label{alg-Kuiper-InvCDF}
\begin{algorithmic}[1]
\Require Probability $x \in[0, 1]$, capacity $n$ of the samples
\Ensure  The value of $\inv{F}_{V_n}(x)$ of the  Kuiper's $V_n$-test.
\Function{KuiperInvCDF}{$x, n$}
\State $ y \gets \ProcName{KuiperUTQ}(1.0 - x, n)$; \quad 	// $ y \gets \ProcName{KuiperLTQ}(x, n)$ is OK.
\State \Return $y$;
\EndFunction
\end{algorithmic}
\end{breakablealgorithm}

\section{Verification and Validation}

\label{sec-V-V}

\subsection{An Error in Kuiper's Numerical Results}

Kuiper \cite{Kuiper1960TestsCR} presented a table of the critical value $c^\alpha_n$ for $\alpha\in \set{0.01, 0.05, 0.1}$ and $n\in \set{10, 20, 30, 40, 100, +\infty}$, see \Tab \ref{tab-Kuiper1965}.

\begin{table}[htbp]
\centering
\caption{Kuiper's critical value $c^\alpha_n$ such that $\alpha = \Pr\set{\sqrt{n}\cdot V_n > c^\alpha_n}$} 
\label{tab-Kuiper1965}
\begin{tabular}{|c|cccccc|}
\hline
\diagbox{$\alpha$}{$c^\alpha_n$}{$n$}  & $10$ & $20$ & $30$ & $40$ & $100$ & $+\infty$ \\
\hline
$0.10$ &    $1.1877$ & $ 1.5322$ & $ 1.5503$ & $ 1.5608$ & $ 1.5839$ & $ 1.6196$  \\
$0.05$ &    $1.6066$ & $ 1.6564$ & $ 1.6760$ & $ 1.6869$ & $ 1.7110$ & $ 1.7473$                                               \\
$0.01$ &   $1.8391$ & $1.9027$  & \textcolor{red}{$\bf{1.9153}$} & $1.9375$ & $1.9637$ & 
$2.0010$  \\
\hline 
\end{tabular}
\end{table}

For the configuration of $\alpha$ and $n$, the Kuiper pair for Kuiper's $V_n$-test can be computed by the algorithms above and code released on the GitHub by the authors.  \Tab \ref{tab-KuiperTest} lists some typical results for the pair $\mpair{c^\alpha_n}{v^\alpha_n}$. Our results coincide with Kuiper's numerical results
very well and the readers can check the values in \Tab \ref{tab-KuiperTest} and their counterparts in \Tab \ref{tab-Kuiper1965}.  It should be noted that extra values for $n\in \set{180, 10^6}$ are added in the table. Moreover, the readers can get more effective digits with our code released on the GitHub.
\begin{table}[htbp]
	\centering
	\caption{Kuiper pair for Kuiper's $V_n$-test where $\alpha=\Pr\set{\sqrt{n}\cdot V_n > c^\alpha_n}=\Pr\set{V_n>v^\alpha_n}$}
	\label{tab-KuiperTest}
	\resizebox{\textwidth}{!}{%
		\begin{tabular}{|c|ccccccc|}
			\hline
			\diagbox{$\alpha$}{$(c^\alpha_n, v^\alpha_n)$}{$n$} & $10$ & $20$  &  $30$  & $40$  & $100$  & $180$   & $10^6$ \\ \hline
			$0.10$ & $(1.4877, 0.4704)$  & $(1.5322, 0.3426)$  & $(1.5503, 0.2830)$  & $(1.5606, 0.2468)$  & $(1.5838, 0.1584)$  & $(1.5934, 0.1188)$ & $(1.6193, 0.0016)$       \\
			$0.05$ & $(1.6066, 0.5080)$  & $(1.6563, 0.3704)$   & $(1.6758, 0.3060)$  & $(1.6868, 0.2667)$  
			& $(1.7110, 0.1711)$  & $(1.7208, 0.1283)$  & $(1.7469, 0.0017)$       \\
			$0.01$ & $(1.8401, 0.5819)$  & $(1.9026, 0.4254)$   & \textcolor{blue}{$\bf{(1.9252, 0.3515)}$}  & $(1.9374, 0.3063)$  & 
			$(1.9636, 0.1964)$  
			& $(1.9739, 0.1471)$  & $(2.0006, 0.0020)$      \\ \hline
		\end{tabular}%
	}
\end{table}

It should be noted that there is an error in Kuiper's table for $(\alpha, n, c^\alpha_n) = (0.01, 30, 1.9153)$. The comparison of \Tab \ref{tab-KuiperTest} and 
\Tab \ref{tab-Kuiper1965} shows that the value $c^{0.01}_{30}$ by Kuiper is wrong with the help of our code released on GitHub. Obviously, the correct value should be $c^{0.01}_{30} = 1.9252$ by our algorithms and code or $c^{0.01}_{30} = 1.9253$ according to Kuiper's approximation in which only the term $(\cdot)\me^{-2c^2}$ is considered for the infinite series.  We deem that this error might be a typo due to the  manual work on editing the data since in the 1960 era it  is difficult for a researcher to find an automatic tool for processing data.

\subsection{Table of Kuiper pair of  $V_n$-test for $n=10$}

Kuiper \cite{Kuiper1960TestsCR} also presented a table of $\mpair{v^\alpha_n}{c^\alpha_n}$ for $n=10$ and $c^\alpha_n \in \set{1.0, 1.1, 1.2, 1.3, 1.4, 1.5, 1.6, 1.7, 1.8, 1.9}$. Our programs also give similar numerical results, as shown in \Tab \ref{tab-c-alpha-n10}.
\begin{table}[htbp]
\centering
\caption{Kuiper pair $\mpair{v^\alpha_n}{c^\alpha_n}$ for $n=10$.}
\label{tab-c-alpha-n10}
\begin{tabular}{|c|c|c|}
\hline
\textbf{Critical value} &   \textbf{Upper Tail Quantile}  &   \textbf{Upper tail probability}  \\
 $c^\alpha_n$ & $v^\alpha_n=c^\alpha_n/\sqrt{n}$ & $\alpha = \Pr\set{\sqrt{n}\cdot V_n > c^\alpha_n}$ \\
\hline 
	 $1.00$ & 		 $0.3163$ & 	 $0.6930$ \\
	 $1.10$ & 		 $0.3482$ & 	 $0.5280$ \\
	 $1.20$ & 		 $0.3795$ & 	 $0.3770$ \\
	 $1.30$ & 		 $0.4110$ &  	 $0.2520$ \\
	 $1.40$ & 		 $0.4427$ & 	 $0.1580$ \\
	 $1.50$ & 		 $0.4746$ & 	 $0.0930$ \\
	 $1.60$ & 		 $0.5060$ & 	 $0.0520$ \\
	 $1.70$ & 		 $0.5382$ & 	 $0.0270$ \\
	 $1.80$ & 		 $0.5692$ & 	 $0.0135$ \\
	 $1.90$ & 		 $0.6006$ & 	 $0.0063$ \\
\hline
\end{tabular}
\end{table}

\subsection{Special case of Sufficiently Large Number of Samples}

If the number of samples is large enough, we can let $n$ approach to infinity. Therefore, 
\begin{equation}
\begin{cases}
A_1(c, \infty) = -2 + 8c^2 \\
A_2(c, \infty) = -2 + 32c^2 \\
\mathcal{A}^\alpha_\infty(f,c) = \sqrt{\frac{\ln\left[-2+8c^2 + (-2+32c^2)\me^{-6c^2}\right] -\ln \alpha}{2}}
\end{cases}
\end{equation}
The Kuiper's critical value is the fixed-point of  
\begin{equation}
c = \mathcal{A}^\alpha_\infty(f,c).
\end{equation}
We start the iterative procedure with initial value $\scrd{c}{guess} = 1.2$ and set $n = 10^{8}$, and provide the $c^\alpha_\infty$ for typical values of $\alpha$ in \Tab \ref{tab-c-alpha-n-infty}, which also coincides with Kuiper's original results for $\alpha \in \set{0.1, 0.05, 0.01}$ very well but our results illustrate more possible values of $\alpha$. 
\begin{table}[htbp]
\caption{Kuiper's critical value for $n\to \infty$ (e.g., we can set $n = 10^{8}$ in computer program)} 
\label{tab-c-alpha-n-infty}
\resizebox{\textwidth}{!}{
\begin{tabular}{|l|cccccccccccc|}
\hline
$\alpha$ &  $0.10$ & $0.09$ & $0.08$ & $0.07$ & $0.06$ & $0.05$ & $0.04$ & $0.03$ & $0.02$ & $10^{-2}$ & $10^{-6}$ & $10^{-10}$\\
\hline 
$c^\alpha_\infty$ & $1.6196$  & $1.6400$ & $1.6623$ & $1.6871$ & $1.7150$ & $1.7472$ & $1.7855$ & $1.8331$
& $1.8974$ & $2.0009$ & $3.0056$ & $3.7226$\\
\hline
\end{tabular}
}
\end{table}

\subsection{Kuiper pair for Kuiper's $V_{n,n}$-test}

\Tab \ref{tab-KuiperTest-Vnn} demonstrates the Kuiper pair for $\alpha \in \set{0.01, 0.02, \cdots, 0.10}$ and $n\in \set{10, 20, 30, 40, 100, 10^8}$ with the help of the procedure \ProcName{KuiperPairSolver} in \Algor \ref{alg-KuiperPairSolver}. We remark that the corresponding values for $\alpha$ proposed
by Kuiper \cite{Kuiper1960TestsCR} is $\alpha\in \set{0.10, 0.05, 0.01}$ and the values of $v^\alpha_{n,n}$ is not listed.

\begin{table}[H]
\centering
\caption{Kuiper pair for Kuiper's $V_{n,n}$-test where $\alpha=\Pr\set{\sqrt{n}\cdot V_{n,n} > c^\alpha_{n,n}}=\Pr\set{V_{n,n}>v^\alpha_{n,n}}$}
	\label{tab-KuiperTest-Vnn}
\resizebox{\textwidth}{!}{
\begin{tabular}{|c|cccccc|}
\hline
\diagbox{$\alpha$}{$(c^\alpha_{n,n}, v^\alpha_{n,n})$}{$n$} & $10$     & $20$      & $30$     & $40$     & $100$  & $10^8$ \\ \hline
$0.10$ & $(2.2431, 0.7093)$ &   $(2.2660, 0.5067)$ &   $(2.2740, 0.4152)$ &   $(2.2780, 0.3602)$ &   $(2.2854, 0.2285)$ &   $(2.2905, 0.0002)$ \\ \hline 
$0.09$ & $(2.2682, 0.7173)$ &   $(2.2929, 0.5127)$ &   $(2.3015, 0.4202)$ &   $(2.3058, 0.3646)$ &   $(2.3139, 0.2314)$ &   $(2.3193, 0.0002)$ \\ \hline 
$0.08$ & $(2.2953, 0.7258)$ &   $(2.3220, 0.5192)$ &   $(2.3314, 0.4257)$ &   $(2.3362, 0.3694)$ &   $(2.3449, 0.2345)$ &   $(2.3509, 0.0002)$ \\ \hline 
$0.07$ & $(2.3248, 0.7352)$ &   $(2.3540, 0.5264)$ &   $(2.3643, 0.4317)$ &   $(2.3696, 0.3747)$ &   $(2.3793, 0.2379)$ &   $(2.3860, 0.0002)$ \\ \hline 
$0.06$ & $(2.3572, 0.7454)$ &   $(2.3896, 0.5343)$ &   $(2.4011, 0.4384)$ &   $(2.4070, 0.3806)$ &   $(2.4180, 0.2418)$ &   $(2.4255, 0.0002)$ \\ \hline 
$0.05$ & $(2.3933, 0.7568)$ &   $(2.4298, 0.5433)$ &   $(2.4430, 0.4460)$ &   $(2.4497, 0.3873)$ &   $(2.4623, 0.2462)$ &   $(2.4710, 0.0002)$ \\ \hline 
$0.04$ & $(2.4343, 0.7698)$ &   $(2.4764, 0.5537)$ &   $(2.4918, 0.4549)$ &   $(2.4998, 0.3952)$ &   $(2.5147, 0.2515)$ &   $(2.5251, 0.0003)$ \\ \hline 
$0.03$ & $(2.4819, 0.7849)$ &   $(2.5321, 0.5662)$ &   $(2.5508, 0.4657)$ &   $(2.5607, 0.4049)$ &   $(2.5793, 0.2579)$ &   $(2.5924, 0.0003)$ \\ \hline 
$0.02$ & $(2.5393, 0.8030)$ &   $(2.6021, 0.5819)$ &   $(2.6266, 0.4796)$ &   $(2.6397, 0.4174)$ &   $(2.6650, 0.2665)$ &   $(2.6834, 0.0003)$ \\ \hline 
$0.01$ & $(2.6124, 0.8261)$ &   $(2.6986, 0.6034)$ &   $(2.7351, 0.4994)$ &   $(2.7556, 0.4357)$ &   $(2.7973, 0.2797)$ &   $(2.8297, 0.0003)$ \\ \hline 
		\end{tabular}
		}
\end{table}

\subsection{Cumulative Distribution Function for Kuiper's $V_n$-test}

With the help of \eqref{eq-inv-CDF} and \Algor \ref{alg-Kuiper-InvCDF}, we can compute the $x = \inv{F}_{V_n}(y)$ by calling the procedure \ProcName{KuiperInvCDF} as follows 
\begin{equation}
 x_i \gets \ProcName{KuiperInvCDF}(y_i, n), \quad i =1, 2, \cdots, \scrd{i}{size}
\end{equation}
Thus for each capacity $n$ of the random samples $\set{X_t: 1\le t\le n}$ and probability sequence 
$$\set{y_i = F_{V_n}(x_i): 1\le i \le \scrd{i}{size}},$$
we can draw the curve of CDF with the pairs $\set{(x_i, y_i): 1\le i \le \scrd{i}{size}}$ such that $y_i = \Pr\set{V_n \le x_i} = F_{V_n}(x_i)$. 

\begin{figure}[htbp]
\centering
\includegraphics[width=10cm]{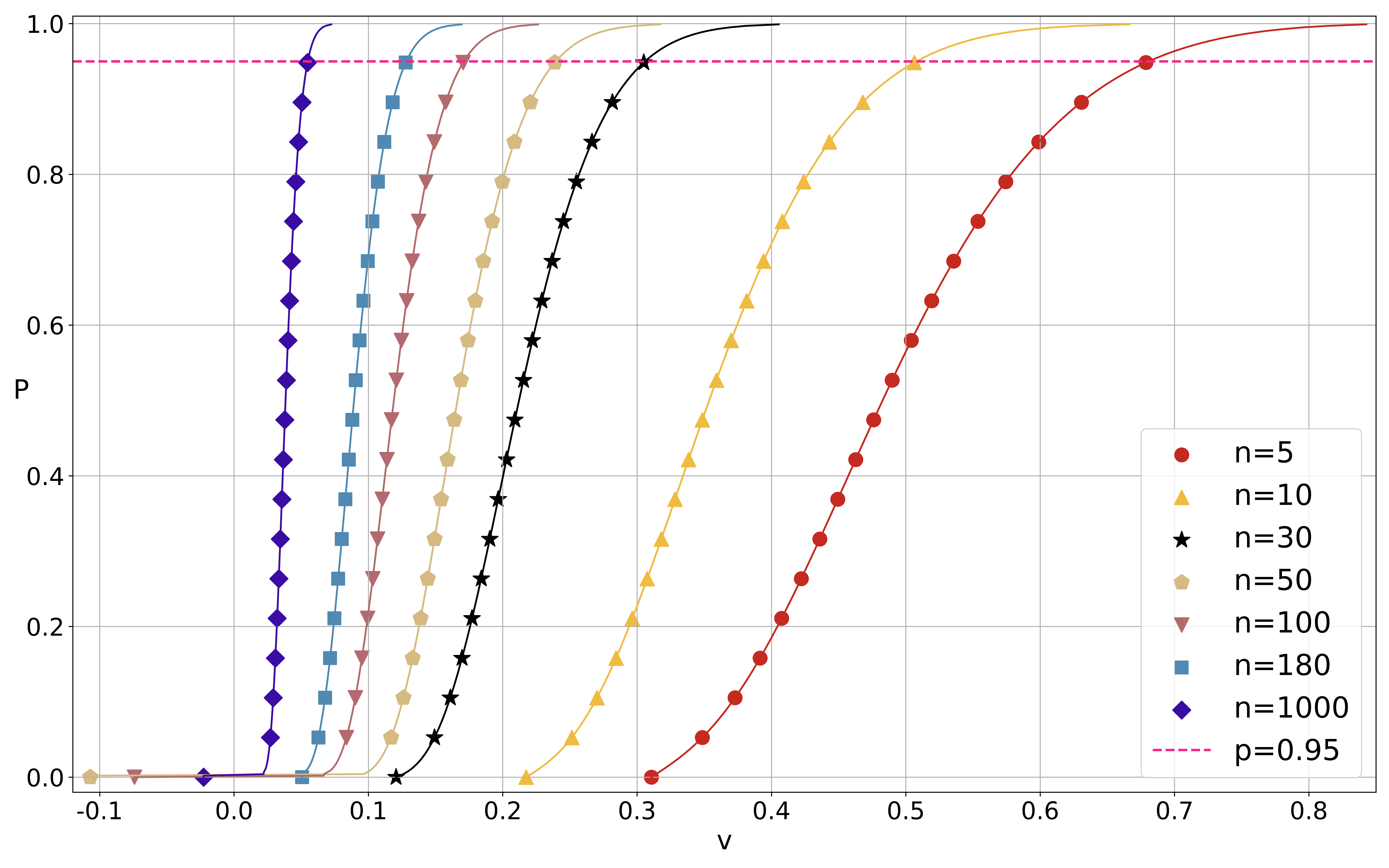} 
\caption{CDF of $V_n$-statistic for various sample capacity $n$} 
\label{fig-Vn-CDF}
\end{figure}

\Fig \ref{fig-Vn-CDF} illustrates the CDF of $V_n$ statistic for various sample capacity $n$. It is obvious that with the increasing of $n$, the curve of CDF becomes more and more steep, which coincides with \eqref{eq-lim-Vn-CDF}. Moreover, the horizontal line $p = F_{V_n}(v)=0.05$ shows the upper tail probability $\alpha = 1-p = 0.05$. The intersection points of the line with the CDF are the upper tail quantile $v^\alpha_n$ for $\alpha = 0.05$ where $n \in \set{5,10,30,50,100,180, 1000}$.
We remark that there are three equivalent ways for computing the inverse of CDF. Actually, we have
\begin{equation}
\begin{aligned}
x &= \inv{F_{V_n}}(y)\\
  &= \ProcName{KuiperInvCDF}(y, n) \\
  &= \ProcName{KuiperLTQ}(y,n) \\
  &= \ProcName{KuiperUTQ}(1-y,n)
\end{aligned}
\end{equation}
for ant $y\in [0,1]$.

\section{Discussion}
\label{sec-discussion}

\subsection{An Alternative Choice of Non-linear Function for $V_n$-test}
\begin{figure}[htbp]
\centering
\includegraphics[width=0.7\textwidth]{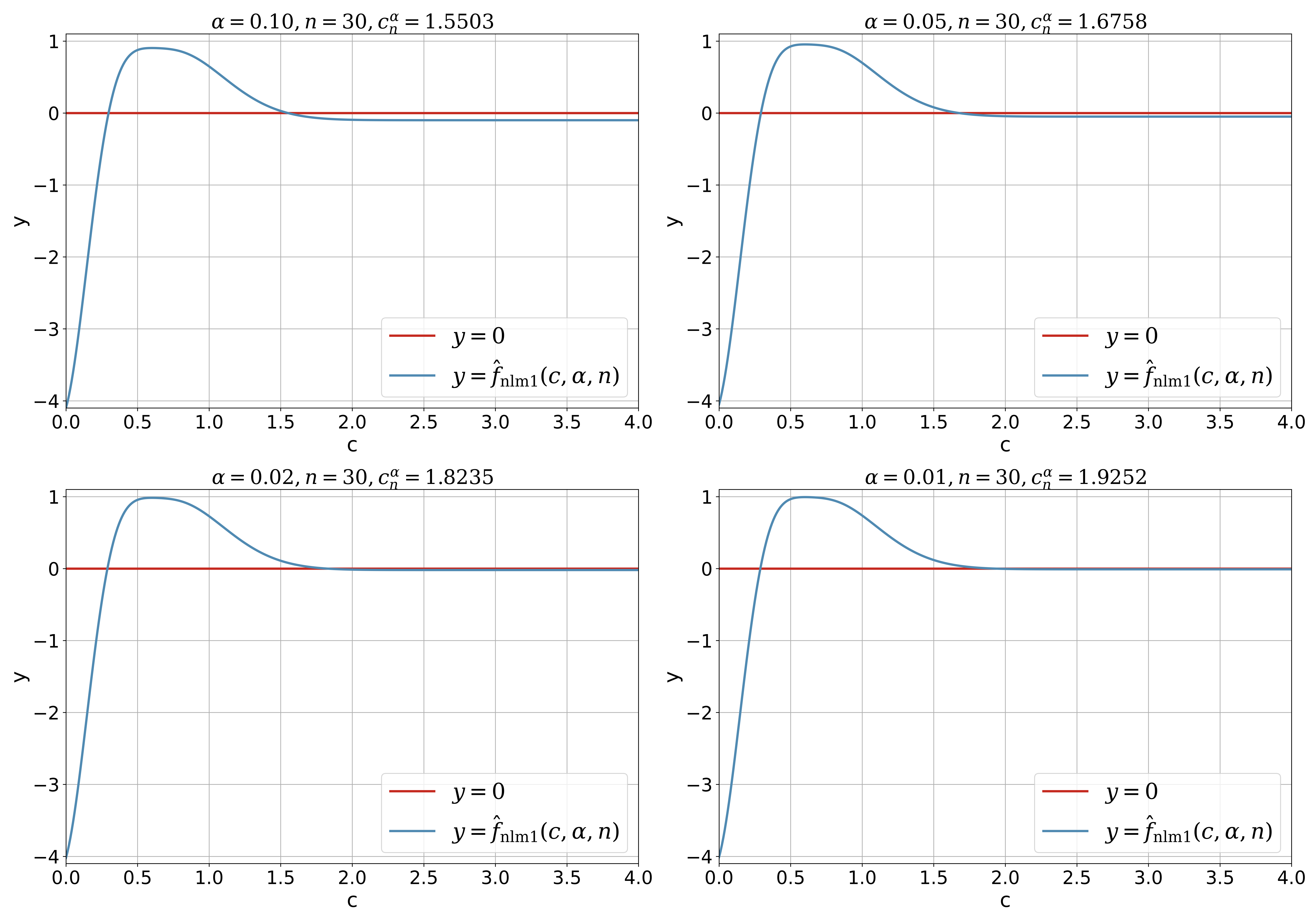} 
\caption{Curve and root of $\scrd{\hat{f}}{nlm1}(c, \alpha, n)$ for $\alpha \in \set{0.10, 0.05, 0.02, 0.01}$ and $n = 30$ in Kuiper's $V_n$-test.}
\label{fig-Qanc-curve-root}
\end{figure}

The disadvantage of Newton's iterative method for solving root lies in two aspects: firstly, derivative of the objective function is required and the method will fail if the derivative is near to 0. Actually, if we take the following alternative equivalent nonlinear equation for $c$
\begin{equation}
 \scrd{\hat{f}}{nlm1}(c, \alpha, n) = A_1(c,n)\cdot \me^{-2c^2} + A_2(c,n)\me^{-8c^2} -\alpha = 0\Longleftrightarrow \scrd{f}{nlm1}(c,\alpha,n) = 0 
\end{equation} 
then for the fixed $\alpha$ and $n$ the function $\scrd{\hat{f}}{nlm1}(c, \alpha, n)$ will have a flat region where the derivative $\scrd{\hat{f}}{nlm}'(c, \alpha, n) =\dif \scrd{\hat{f}}{nlm1}(c,\alpha, n)/\dif c$ will be near zero, thus the Newton's iterative method may fail if the initial value $\scrd{c}{guess}$ is larger than the root. \Fig \ref{fig-Qanc-curve-root} demonstrates this situation clearly in which for $c > c^\alpha_n$ the curve of $\scrd{\hat{f}}{nlm1}(c, \alpha, n)$ is very flat.

\subsection{Choices of Initial Value  for the Critical Value}

Theoretical and numerical experiments indicate the way for setting the initial value for solving the Kuiper pair. Our suggestions are listed as follows:
\begin{itemize}
\item for the $V_n$-test, we can set:   $\scrd{c}{guess}\in (0.5, 2.5)$ for the direct iterative method based on $\mathcal{A}^\alpha_n$  and  $\scrd{c}{guess}\in (1.1, 2.5)$ for the Newton's iterative method based on $\mathcal{B}^\alpha_n$;
\item fort the $V_{n,n}$-test, we can set:  
 $\scrd{c}{guess}\in (2.4, 2.6)$ for the direct iterative method based on $\mathcal{A}^\alpha_{n,n}$  and
 $\scrd{c}{guess}\in (2.2, 2.6)$ for the Newton's iterative method based on $\mathcal{B}^\alpha_{n,n}$.
\end{itemize}
In summary, if we set $\scrd{c}{guess}\in [2.4, 2.5]$, then the algorithms proposed could work well for both $V_n$-test and $V_{n,n}$-test.

\subsection{Limitation of Kuiper's First Order Approximation for the Cumulative Distribution Function}

The limitation of Kuiper's first order approximation for the cumulative distribution function is that the approximation error is bounded by $\BigO{n^{-1}}$. Thus for small $n$, the numerical precision is not high.
Although Stephens \cite{Stephens1970} proposed the modified statistic $T_n = (\sqrt{n} + 0.155 + 0.24/\sqrt{n})V_n$ for replacing the $V_n$ for small $n$, the bound of approximation error is not discussed. Essentially, what we need is a more precise formula for \eqref{eq-utp-inf-series} with smaller approximation error, say $\BigO{n^{-2}}$ or $\BigO{n^{-3}}$, instead of using $T_n$ directly without verification and validation.

\section{Conclusion}
\label{sec-conclusion}

The computation of the critical value and upper tail quantile, Kuiper pair for simplification, is equivalent to solve the fixed-point of the nonlinear equation $\alpha = \Pr\set{K_n > c} = \Pr\set{\sqrt{n}V_n>c}$ which involves two infinite series by iterative method for Kuiper's statistic $V_n$ or
$V_{n,n}$. The Kuiper's pair $\mpair{c^\alpha_\star}{v^\alpha_\star}$ can be solved with the following steps:
\begin{itemize}
\item[(1)] simplifying the nonlinear equation $\alpha = \Pr\set{\sqrt{n}\cdot V_\star>c}$  with second order approximation for $V_\star \in \set{V_n, V_{n,n}}$ in the asymptotic expansion;
\item[(2)] converting the nonlinear equation to a fixed-point equation with the form $c = T(f, c, \alpha, n)$ by setting the updating operator $T$, the function object $f(c, \alpha, n)$ and initial value $\scrd{c}{guess}$ properly via the direct iterative scheme $c_{i+1} = \mathcal{A}^\alpha_\star(c) = f(c_i, \alpha, n)$ or the Newton's scheme $c_{i+1} = \mathcal{B}^\alpha_\star(f, c) = c_i -f(c_i, \alpha, n)/f'_c(c_i, \alpha, n)$;
\item[(3)] designing algorithms for solving the fixed-point equation in order to compute the critical value $c^\alpha_\star$ and the upper tail quantile $v^\alpha_\star$;
\end{itemize}

For the convenience of implementation with concrete computer programming languages such as C, C++, Python, Octave, MATLAB and so on, the pseudo-code for the algorithms are provided with details. There are some advantages for our methods and algorithms:
\begin{itemize}
\item[(1)] a unified theoretic framework for solving fixed-point, based on  the concept of functional in mathematics and the high order function in computer science, is discussed in detail by combining the merits of mathematics and computer science, which is general and elastic enough for solving lots of fixed-point problems arising in science, technology, engineering and mathematics;
\item[(2)] a unified interface is set for  solving the Kuiper's pairs in the $V_n$-test and $V_{n,n}$-test;
\item[(3)] procedures for solving the upper/lower tail quantile are provided for the potential applications of the Kuiper's statistic in the goodness-of-fit test;
\item[(4)] the computational complexity is linear since there is no nested loops in the algorithm, and the reader can test the running time with the help of the C code released on GitHub;
\item[(5)] the methods proposed in this paper can be modified slightly to solve the Kolmogrov-Smirmov test, $\chi^2$-test and normal test since the difference behind these different tests lies in the concrete form of the nonlinear equation $\alpha = \Pr\set{Z >z}$ for the given upper tail significance level $\alpha$ and the CDF $F_Z(\cdot)$ encountered for the population $Z$ and random samples $\set{Z_t: 1\le t\le n}$.  
\end{itemize} 
Our verification and validation shows that there is a mistake in Kuiper's table of critical value $c^\alpha_n$ for $(\alpha, n) = (0.01, 30)$. The correct value for $c^{0.01}_{30}$ should be $1.9252$ or $1.9253$ instead of $1.9153$. It might be a typo introduced by manual work on editing the data in the 1960 era. 

In the sense of STEM education, the topic of this paper can be used as a comprehensive project for training the college students' ability of solving complex problems by combining the mathematics and computer programming.

\subsubsection*{Acknowledgements} 
This work was supported in part by the Hainan Provincial Natural Science Foundation of China under grant numbers 720RC616 and 2019RC199, and in part by the National Natural Science Foundation of China under grant number 62167003.

\subsection*{Code availability statement}

The code for the implementations of the algorithms discussed in this paper can be downloaded from the following GitHub website: \textcolor{blue}{\url{https://github.com/GrAbsRD/KuiperVnStatistic}}.

\end{document}